\documentclass[10pt,letterpaper]{article}
\usepackage[utf8]{inputenc}
\usepackage[T1]{fontenc}
\usepackage{textcomp}
\usepackage{mathptmx}
\usepackage[margin=1in]{geometry}
\usepackage{microtype}
\usepackage{graphicx}
\usepackage{array,booktabs,longtable}
\usepackage[labelfont=bf,labelsep=colon,font=small,skip=6pt]{caption}
\usepackage{enumitem}
\setlist[itemize]{itemsep=2pt,topsep=4pt}
\usepackage[hidelinks,pdfusetitle]{hyperref}
\usepackage{xurl}
\DeclareUnicodeCharacter{2212}{\textminus}
\DeclareUnicodeCharacter{00B7}{\textperiodcentered}
\title{%
  \texorpdfstring{\vspace*{-2.3em}}{}
  \textbf{AI Agent Observability in Finance:\\
    A Survey of LLM Trading Agents}%
}
\newcommand{\affmark}[1]{\texorpdfstring{\textsuperscript{#1}}{}}
\newcommand{\authorsep}{\texorpdfstring{\qquad}{\and}}
\newcommand{\emailaddress}[1]{\href{mailto:#1}{\nolinkurl{#1}}}
\newcommand{\authoremail}[2]{\affmark{#1}\,\emailaddress{#2}}
\author{%
  Yuhan Ren\affmark{1,}\thanks{%
    \urlstyle{same}\textit{Corresponding author}\enspace
    Yuhan Ren (\emailaddress{leonren@alumni.upenn.edu})%
  }\authorsep
  Yunbai Zhang\affmark{2}\authorsep
  Nan Jiang\affmark{3}\authorsep
  Heze Ma\affmark{4}%
  \texorpdfstring{\\[.25em]
    {\normalsize\urlstyle{same}%
      \begin{tabular}{@{}c@{}}
        \affmark{1}\,University of Pennsylvania, Philadelphia, Pennsylvania, USA\\
        \affmark{2,4}\,Columbia University, New York, New York, USA\\
        \affmark{3}\,University of Miami, Coral Gables, Florida, USA\\[.2em]
        \small\authoremail{2}{yz3386@columbia.edu}\quad
        \authoremail{3}{nxj286@miami.edu}\quad
        \authoremail{4}{hm2938@columbia.edu}%
      \end{tabular}}%
  }{}%
}
\date{\small November 2026}

\begin{document}
\maketitle
\enlargethispage{.5\baselineskip}
\vspace{-2\baselineskip}
\begin{abstract}
Large language model (LLM) agents now read market data, write investment theses, size positions and place orders, some with real money. A firm that deploys one must answer the questions it asks of any model: what the agent knew, why it acted, what it did, and how far that account can be trusted. The agent's own explanation is a weak answer, because stated reasoning can diverge from the process behind a trade and stated confidence is miscalibrated. We treat the problem as one of observability: capturing and analyzing agent behavior in a form that risk, compliance and audit teams can use as evidence. Our taxonomy, FinAO (Finance Agent Oversight), classifies recent work by what it observes, the risk it controls, when and by whom it is applied, and where its evidence is recorded and how far that evidence can be trusted. Where trading-specific evidence is limited, we draw on agent observability research and supervisory texts. The literature tests data inputs and research signals thoroughly, but it rarely observes an agent's books and records, and almost none of it addresses audit, incident review or supervisory examination. No study records everything needed to reconstruct a single decision. We map these gaps to supervisory expectations and propose an audit sub-agent that works only from records it cannot change. Three questions remain open. How can a firm reconstruct a single trading decision from records rather than from the agent's narrative? After an incident, how can a reviewer find every later decision a corrupted input affected? How can a supervisor examine many agents at once without relying on any one agent's account?

\noindent\textbf{Keywords:} LLM agents, algorithmic trading, model risk management, observability, AI observability, telemetry, audit, financial supervision
\end{abstract}

\section{Introduction}

LLM trading agents matured quickly in recent years. Early systems such as FinMem [122] paired a single model with layered memory. Later systems copied the structure of a trading firm, with analyst, researcher, trader and risk roles that debate before acting [68, 112]. Reinforcement learning now tunes models directly on trading rewards [51, 113, 115], and live arenas let agents trade real or paper money in public [25, 89, 98]. A firm that deploys such an agent takes on its decisions. When a trade goes wrong, the firm's risk, compliance and audit functions must reconstruct what the agent knew, why it acted and whether the action was authorized. A supervisor may later ask the same questions.

In finance, a decision must be traceable, by someone other than the decision-maker, to the evidence available when it was made. Model risk management supplies the vocabulary. A model is developed, independently validated, monitored in use and audited, with responsibilities split across three lines of defense. Agents strain this framework in ways that conventional algorithms and models do not (Section 2.1), and the newest US guidance leaves agentic AI outside its scope. Model Risk Management for Generative AI in Financial Institutions [7] extends the framework to generative models. Agent Reliability Profiles [43] proposes assurance levels for agent deployments in financial services, and Agentic AI Systems and Financial Stability [86] models how single-agent model risk becomes systemic. Both assume that an agent's behavior can already be turned into evidence that a validator, an auditor or a supervisor can rely on, and neither maps the trading-agent literature.

Model risk management thus says who must rely on evidence about an agent, but not how the agent's behavior becomes evidence. We call that capability AI agent observability. In financial services, it is the capture and analysis of an agent's behavior, including its data access, tool calls, outcomes and the human approvals and overrides around it, in a form that risk, compliance and audit teams can use as evidence. Telemetry is the raw material: traces, logs and metrics. Observability is the ability to answer questions about the agent from that material. A simple test applies: from the records alone, can a reviewer reconstruct a single decision, including its inputs, the context it retrieved, its tool calls, its output and any human intervention?

Agent observability research supplies the capture side, but it was built for engineers who operate agents, not for the independent review that model risk management requires. AgentOps [21, 110] defines what to trace and frames agent operations as monitoring, anomaly detection, root-cause analysis and resolution. That covers monitoring in use, but not independent validation or audit. The OpenTelemetry GenAI conventions [94] standardize agent and tool spans but carry no finance semantics, such as the as-of time of an input or the intent behind an order. Provenance work [109] supplies the typed links that FinAO's record model adopts. Diagnosis, which validators and auditors need most, lags capture. On the Who\&When benchmark, the best method found the responsible agent in 53.5 percent of failed multi-agent runs but the decisive step in only 14.2 percent [126], and telemetry shaped by current conventions supports detecting a failure far better than locating where it began [130]. Intervention-based debugging [82], receiver-signed receipts [28] and claim-to-evidence graphs [54] point to evidence strong enough for independent review, while reasoning traces remain useful but fragile [57]. None of this work addresses trading, market data or the lines of defense.

We organize the literature around the questions a firm's control functions ask. What does a method observe, and at which stage of the investment process? Which risk does it control? When, and by whom, is it applied? Where is its evidence recorded, and how far can it be trusted? We exclude pre-trade limits, kill switches and other runtime controls. They consume the same records, but this survey studies the evidence itself and the after-the-fact review that relies on it. To our knowledge, this is the first survey to treat LLM trading agents as models under model risk management and to ask what must be recorded for them to be audited.

Existing surveys of trading agents ask how agents are built and how well they perform. Large Language Model Agent in Financial Trading [18] reviews early architectures and their backtests. Agentic Trading [111] and Agentic Quantitative Trading [46] map architectures and workflows, From Algorithms to Agents [27] links capabilities to supervisory frameworks, and Execution Realism and Reproducibility [119] and FARSIGHT [107] audit reproducibility and robustness. Broader surveys cover LLMs in financial prediction and trading [31], financial AI more generally [74] and multi-agent market simulation [75], and Toward Reliable Evaluation [88] proposes a taxonomy for evaluating financial multi-agent systems. None asks what must be recorded to review an agent.

The survey makes four contributions.

\begin{itemize}
\item \textbf{Observability taxonomy.} FinAO classifies each study by investment-process stage, risk type, model-risk lifecycle phase and evidence quality, and fixes the signals that must be recorded for an agent's behavior to be traceable.
\item \textbf{Evidence assessment.} We grade each study's evidence on an audit-style hierarchy and show where independent evidence is missing.
\item \textbf{Signal coding.} We code which observability signals each study records, and find that none records everything needed to reconstruct a decision.
\item \textbf{Challenges and an audit architecture.} We identify the judgments that remain difficult when a firm or supervisor reviews agents after the fact, including multi-agent systems, and propose an audit sub-agent for them.
\end{itemize}

Throughout, we follow one fictional trade in NVIDIA (NVDA) stock; the headline, trades and records are invented for illustration. On Day 1 an altered headline, “NVDA upgraded”, reaches a trading agent's news feed, and the agent stores the belief that NVDA's outlook is positive. It buys 2,000 shares on Day 3 and 1,000 more on Day 5, and on Day 9 a loss starts a review. Figure 2 contrasts the agent's account of this trade with independent records, Table 1 shows what each observability signal would reveal about it, Section 6.1 and Figure 4 write out the record of the Day 5 order, and Figure 5 traces the review.

The reviewed studies show that look-ahead bias, miscalibration and unfaithful explanations can be measured and partly corrected. It remains much harder to reconstruct a specific decision, trace an incident through an agent's memory, or examine many agents at once. Section 2 explains what agents change for oversight and introduces the taxonomy, Section 3 reviews methods by investment-process stage, Section 4 examines what current evidence establishes, Section 5 maps it to supervisory expectations, and Section 6 discusses the remaining challenges.

\section{Agents and the FinAO Taxonomy}

We classify recent studies on trading agents, alpha mining, models tuned by reinforcement learning, benchmarks, look-ahead bias, calibration, interpretability, adversarial robustness, market simulation, collusion and model-risk governance. Some studies could be read only as abstracts, summaries or citing papers, and we treat their classifications as provisional. Related surveys, agent observability research and supervisory texts inform the analysis but are not classified.

\subsection{What Is New About Agents}

Firms already oversee algorithms and models. MiFID II RTS 6 [22] requires investment firms to test trading algorithms before deployment, apply pre-trade controls and real-time monitoring, and keep records of their activity. SR 11-7 [8] set the US model risk standard for more than a decade, with validation built on conceptual soundness, ongoing monitoring and outcomes analysis by staff independent of the developers. Broker-dealer recordkeeping rules [105] and order audit trails let a firm or supervisor reconstruct what was ordered, when and by whom. These regimes rest on three assumptions. The logic behind a decision is written down and can be inspected. The system's state is the firm's book, which is reconciled every day. A validated version, given the same inputs, does the same thing until a controlled change replaces it.

LLM agents break each assumption. The logic is not written down: the agent supplies a free-text rationale, and rationales can misstate why the agent acted [101] or which inputs mattered [2]. Agents keep their own state, in memory and beliefs that persist across sessions, shape later orders and can be corrupted without tripping risk controls [117]. Their decisions are not reliably repeatable on identical inputs [53]. The model can also change underneath the agent. Hosted models are updated by vendors, so the validated configuration may not be the one running [27, Sec. 9.4], and a model's actual knowledge cutoff can differ from the one its vendor states [50].

Supervisory texts reflect the gap. SR 26-2 [9] and OCC Bulletin 2026-13 [90], which replaced SR 11-7 in April 2026, describe generative and agentic AI as novel and rapidly evolving and place them outside their scope. ESMA's briefing on algorithmic trading [23] keeps the firm fully responsible for third-party algorithms and extends governance and explainability expectations to AI. The order audit trail still shows what an agent ordered, but no longer why. That evidence must come from records of the agent's inputs, state, configuration and authorization, which the rest of this survey examines.

\subsection{Four Questions}

FinAO treats an LLM trading agent as a model under model risk management whose behavior must be observable: traceable through metrics, logs and traces, and understandable to the people who rely on it. It asks four questions of any oversight method, and each study becomes a tuple of answers, with possibly several values for one question (Figure 1). Two categories hold the fewest reviewed studies: books and records among the stages, and audit, incident review and supervisory examination among the lifecycle phases (Sections 3.3 and 4.2). Appendix A defines the record model and the signals each question calls for.

\begin{figure}[htbp]
\centering
\includegraphics[width=0.82\linewidth]{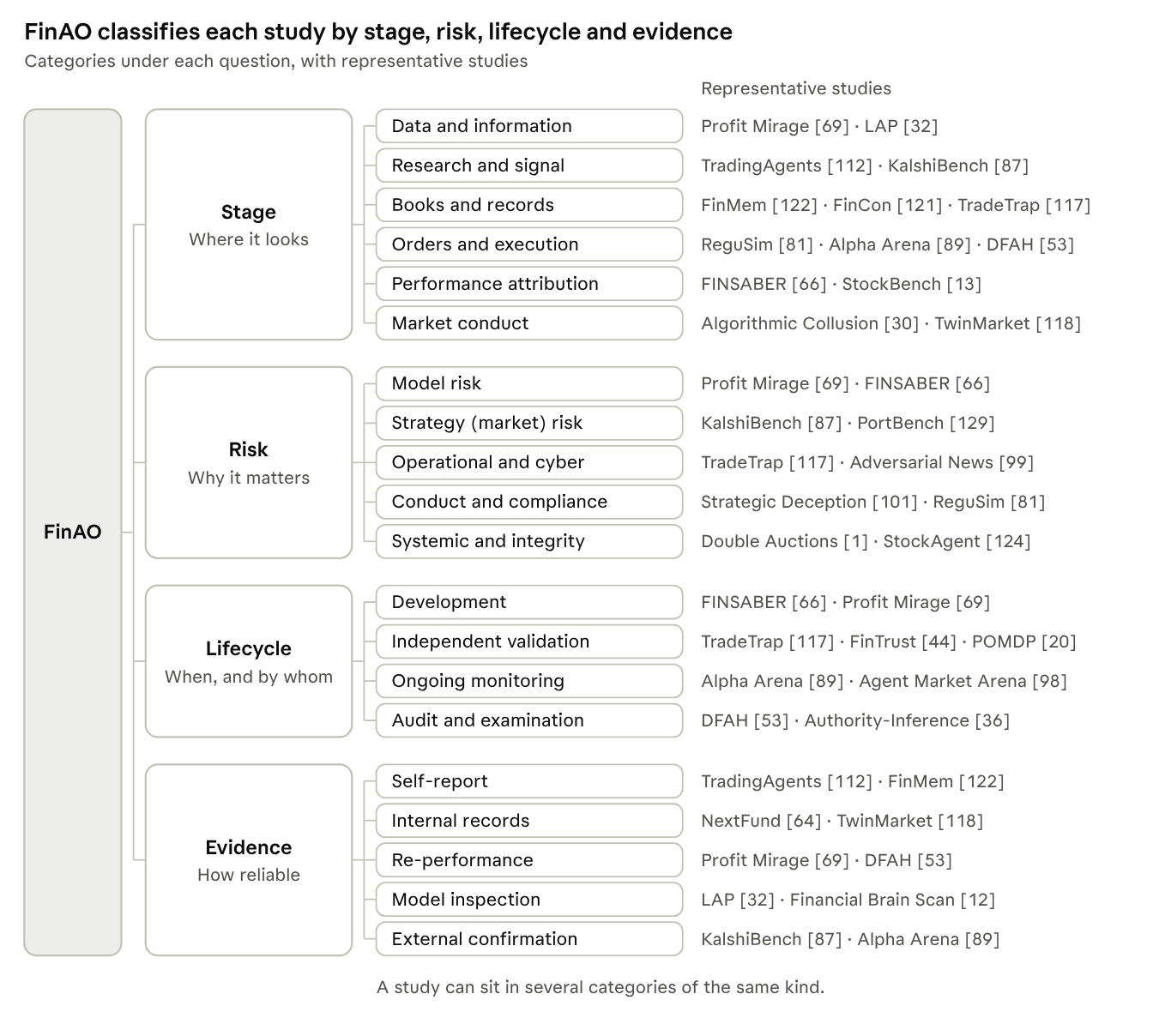}
\caption{The FinAO taxonomy: investment-process stage, risk type, lifecycle phase and evidence quality, with their categories and representative studies. Bracketed numbers refer to the references.}
\label{fig:fig1}
\end{figure}

\textbf{Investment-process stage} says where a method looks. The six stages are data and information, research and signal generation, books and records, orders and execution, performance and risk attribution, and market conduct and stability. They form a loop: data feeds research, research drives orders, outcomes flow back into the agent's books, and those books shape the next decision. Three targets cut across every stage: runtime, configuration (including which model served which task) and human intervention.

\textbf{Risk type} names the risk a method controls, and with it the function that must act on a finding. The five types are model risk, strategy risk, operational and cyber risk, conduct and compliance risk, and systemic and market-integrity risk. They serve as the taxonomy's failure classes: model risk asks whether an edge is real or an artifact of leakage, overfitting or bias; conduct risk asks whether the firm can attest to why a trade was made; and systemic risk asks whether agents herd, collude or fail together.

\textbf{Lifecycle phase} records when a method applies and which line of defense applies it: development and backtesting (first line), independent validation and stress testing (second line), ongoing monitoring (shared, and the keeper of evaluation history), and audit, incident review and supervisory examination (third line and supervisors). Along the lifecycle, the unit of analysis widens from one decision to a desk, a firm's agents and the market. The task shifts from detecting failures to reconstructing and explaining them.

\textbf{Evidence quality} ranks a method's evidence as an auditor ranks audit evidence. Three properties set the rank: source (declared, observed or inferred), support (observation, a constraint check or a test by intervention) and integrity (self-reported, independently observed or tamper-evident, which includes how records are retained and who can change them). The levels are management representation (the agent's own reasoning and confidence); internal records (order, tool-call and memory logs); re-performance and stress testing; model inspection; and external confirmation from exchange, broker or on-chain records. Evidence from sources outside the agent ranks above the agent's account of itself.

A test of lookahead propensity [32], for example, is classified under data and information, model risk, development and model inspection: it watches data inputs, for model risk, during development, using model-internal evidence. FinAO also implies a design rule. No stage should be signed off on the agent's own account alone, and every stage an agent touches should have at least one source of evidence the agent does not control: re-performance, model inspection or external confirmation.

The four questions meet when a failure crosses stages. In the running example, the altered headline enters with the data, is stored in the agent's books as a belief, shapes the Day 3 thesis during research and drives the Day 5 order. Detecting it is a validation or monitoring task. Explaining afterwards how it spread, and which later decisions it touched, is an audit task that needs records from every stage it passed through. Figure 2 follows the trade through these four stages. In each row, the left panel accepts the agent's own account and the right panel checks it against records the agent cannot change.

\begin{figure}[htbp]
\centering
\includegraphics[width=0.82\linewidth]{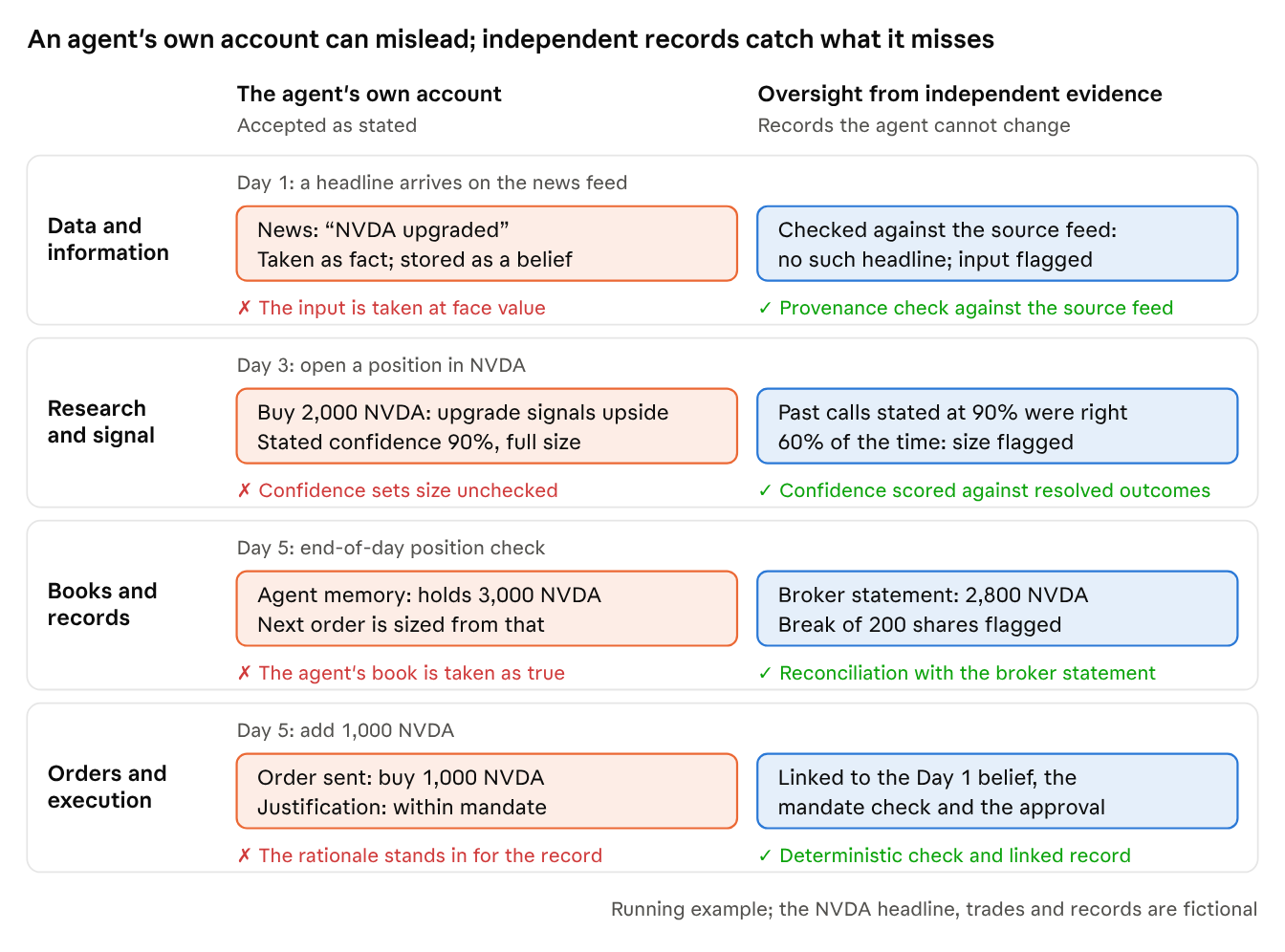}
\caption{The running example at four investment-process stages. The agent's own account versus independent oversight; fictional records.}
\label{fig:fig2}
\end{figure}

\subsection{What Must Be Recorded}

Each question fixes signals that must be recorded for a review to answer it. Table 1 lists ten such signals, what each shows a reviewer, what it would show in the running example, and representative studies that record it. Appendix A gives the record model that links the signals, the metrics and logs for each stage (Table A1), and the rule we used to code each signal (Table A2). An agent's own explanation is weak evidence (Section 4), but a human-readable account can still be built from these records, provided every statement in it cites the records it rests on [54]. That is the rule the audit sub-agent in Section 6.5 follows.

\begingroup\footnotesize\setlength{\tabcolsep}{4pt}\renewcommand{\arraystretch}{1.18}
\begin{longtable}{>{\raggedright\arraybackslash}p{0.126\dimexpr\linewidth-8\tabcolsep\relax}>{\raggedright\arraybackslash}p{0.383\dimexpr\linewidth-8\tabcolsep\relax}>{\raggedright\arraybackslash}p{0.362\dimexpr\linewidth-8\tabcolsep\relax}>{\raggedright\arraybackslash}p{0.114\dimexpr\linewidth-8\tabcolsep\relax}}
\caption{Observability signals, ordered from most to least commonly recorded in the reviewed studies.}\\
\toprule
\textbf{Signal} & \textbf{What it shows a reviewer} & \textbf{In the running example} & \textbf{Representative studies} \\
\midrule
\endfirsthead
\toprule
\textbf{Signal} & \textbf{What it shows a reviewer} & \textbf{In the running example} & \textbf{Representative studies} \\
\midrule
\endhead
\bottomrule
\endlastfoot
Reasoning and confidence & Why the agent says it acted and how sure it claims to be: a claim to test, not a record & The Day 3 thesis “upgrade signals upside”, stated at 90\% confidence & [89, 101, 112, 121] \\
Time-stamped inputs & Which inputs a decision used, where each came from and whether it was available at the time & The Day 1 headline with its feed and arrival time, which the source feed does not contain & [32, 50, 63, 69] \\
Model version and configuration & Which model, prompt and parameters were running, so that a replay divergence can be attributed & The model and prompt versions behind the Day 3 and Day 5 orders & [32, 40, 41, 50] \\
Messages between agents & Which agent passed which belief or task to which & The analyst agent's message passing the NVDA belief to the trader agent & [1, 85, 112, 121] \\
Evaluation over time & Whether stated confidence and performance hold up as outcomes arrive & Past calls stated at 90\% were right 60\% of the time & [25, 64, 89, 98] \\
State and memory & When a belief or position entered the agent's memory, and which later decisions read it & The belief “NVDA outlook positive”, written on Day 1 and read on Days 3 and 5 & [20, 117, 121, 122] \\
Order linked to intent & The rationale, intent, mandate check and authorization behind each order & The Day 5 order, linked to the Day 1 belief and its mandate check & [36, 64, 81, 89, 101] \\
Human intervention & Who approved, overrode or escalated which order, and why & The approval that released the Day 5 order & [36, 93, 104, 108, 123] \\
External records & Exchange, broker, on-chain or resolved-market records outside the agent's control & The broker statement of 2,800 shares against the agent's 3,000 & [36, 72, 87, 89, 98] \\
Identical-input replay & Whether the decision repeats on the same inputs & The Day 5 decision rerun with and without the headline & [33, 36, 53] \\
\end{longtable}\endgroup

\section{Oversight Across the Investment Process}

This section reviews methods by the investment-process stage they observe. Within each stage, we separate studies that build agents from studies that test them, because the two produce different kinds of evidence. A study that builds an agent usually reports the agent's own reasoning and a backtest. A study that tests an agent reruns it, perturbs it or compares it with records it does not control. Each stage closes with what the studies at that stage record.

\subsection{Data and Information}

Data oversight asks whether an agent's inputs were available at the time of the decision and whether they were genuine. For an LLM agent, the first question extends beyond the data feed to the model itself. A model trained after the backtest period may already know what happened. This is the best-covered stage, and most of its studies test the model with evidence it does not control, such as post-cutoff data [69] or anonymized headlines [35].

\subsubsection{Look-ahead and Memorization}

Glasserman and Lin [35] compared GPT-3.5 sentiment on original headlines with versions that removed the company name. In-sample, the anonymized headlines produced better returns. The model's general knowledge of the company distracted it more than look-ahead helped it. Lookahead Bias in Pretrained Language Models [100] used events no one could have predicted at the time. A model named COVID-19 as a risk when reading earnings calls from 2019. Prompting it to ignore later knowledge did not help, while models trained only on earlier text did. The Memorization Problem [80] found that models recall exact values of macro indicators, index levels and returns before their training cutoff and none after. Instructions did not stop the recall, and models rebuilt masked firm names and dates from small clues. Total Recall? [15] found that on a typical day, models believe they hold macroeconomic data that has not yet been released.

Other studies measure the bias directly. A Test of Lookahead Bias [32] scores each firm-date by how likely the model saw the outcome in training. Forecasts were more predictive where that score was high, and the effect vanished after the training cutoff. HindsightBench [50] audits models from the outside by showing, hiding or shifting the date on monthly macro snapshots. All six 2026-generation models tested showed date-triggered hindsight. Actual knowledge cutoffs ran three to six months earlier than vendors stated for the 2026 models, and up to eight months earlier for some 2025 models. Look-Ahead-Bench [6] turns such tests into a standardized benchmark for point-in-time LLMs.

Remedies remove the knowledge or correct for it. ChronoBERT [40] and ChronoGPT-Instruct [41] train a separate model for each year on text published by then. ChronoGPT-Instruct's long-short Sharpe ratio of 0.95, against 1.53 and 1.76 for two unrestricted models, implies that at least 54 to 62 percent of the apparent predictability is genuine. FinCAD [67] corrects at inference time by subtracting what a prompt makes the model recall, cutting inflated in-sample returns by up to 67.1 percent. The size of the bias depends on the task. ChronoBERT found it modest for next-day returns from news [40]. Profit Mirage [69] found that the Sharpe ratios of LLM trading agents fell by 51 to 62 percent after the training cutoff. For time-series foundation models, training on future data did not reliably improve forecasts [11].

\subsubsection{Manipulated Inputs}

A genuine data feed can still carry manipulated content. Adversarial News [99] altered headlines on a single day with look-alike Unicode letters and hidden text. Annual returns fell by 3.67 points on average and by up to 17.70 points, and FinBERT was fooled 67 and 99 percent of the time, depending on the manipulation. AutoRedTrader [77] generated finance-specific fake news and refined it against the agent's reactions. Of the fake items, 69 percent reached the agent's decision and 26.7 percent changed the trade. In Market Signal Injection [61], a rival changed only how market data was presented, never the numbers, and pricing still shifted. Probes on model internals could distinguish attacked runs, but that alone did not flag harmful prices.

In record terms, studies at this stage control which data were available at each decision date, but they do not keep a per-decision record of the inputs a decision used and where each came from. Only NextFund [64], reviewed under attribution, logs the path from evidence to trade. Without such a record, an auditor cannot show that a given decision used only data available at the time.

\subsection{Research and Signal Generation}

Research is where the agent forms its thesis, its signal and its confidence. Most agent-building work sits here, and so does most self-reported evidence: many agent designs report only the agent's own reasoning and a backtest [68, 112, 121]. Large Language Model Agent in Financial Trading [18, Sec. 2] sorts these designs into agents that trade directly and agents that mine alpha factors for a downstream trading system, and notes that every agent it reviewed takes textual data, such as news and filings, as input [18, Sec. 3.2]. Benchmarks such as InvestorBench [65] score these designs on fixed sets of financial decision tasks.

\subsubsection{Agent Designs}

FinMem [122] gives a single agent short-, mid- and long-term memory that fades at different speeds. FinAgent [127] adds charts, tools and two levels of reflection. Later designs copy the structure of a firm. TradingAgents [112] assigns analysts, bull and bear researchers, a trader and a risk team. HedgeAgents [68] coordinates asset-class experts through meetings, and AlphaAgents [128] has agents debate until they agree on stock picks. FinCon [121] updates its investment beliefs by comparing more and less profitable episodes and sends them back to the analysts. Toward Expert Investment Teams [83] found that detailed, analyst-style task prompts raised Sharpe ratios at most portfolio sizes.

Reinforcement learning tunes the model itself. Trading-R1 [113] trains a model to write structured investment theses and turn them into trades. Alpha-R1 [51] uses realized portfolio returns as its reward when deciding which factors to activate. FLAG-Trader [115] uses a partly fine-tuned 135M-parameter model as the trading policy. Fin-R1 [76] rewards correct, well-structured answers to financial questions. In these models, the reward shapes the reasoning trace, so the trace is not a record of how the decision was made.

Alpha-mining agents write signals as formulas. Alpha-GPT [108] turns a researcher's idea into factors through rounds of generation, search and human feedback. Alpha-GPT 2.0 [123] extends the loop to modeling and analysis, and QuantaAlpha [38] evolves factor-mining runs. A formula can be backtested and inspected, which makes these systems easier to review than free-text decisions.

The evidence these studies report is mostly the agent's own: debate transcripts, reflections and backtests. Several test windows fall inside the model's training data. QuantAgents [70] tested on 2021 to 2023, before its model's cutoff, and FinMem's return fell by about 72 percent when rerun past its model's cutoff [69].

\subsubsection{Calibration}

An agent's stated confidence often sets its position size, so calibration is a research-stage risk with consequences at execution. KalshiBench [87] scored frontier models on prediction-market questions that resolved after their training cutoffs. Every model was overconfident, extra reasoning made calibration worse, and only one model beat the base rate. FinBench [34] proposes time-gated forecasts scored with proper scoring rules. Its pilot covers one day and three stocks, so it establishes a design rather than evidence about losses. FinAbstain [104] lets an agent abstain when evidence is thin, which raised accuracy on the forecasts it did make. In Alpha Arena [89], stated confidence did not track performance.

\subsubsection{Bias and Faithfulness}

Models bring their own views. Your AI, Not Your View [62] found that most models prefer technology stocks, large caps and contrarian bets, and cling to first judgments as counter-evidence accumulates. Fin-Bias [45] found that models largely copy whatever analyst rating appears in the report, even a fake one. InvestAlign [106] found that untuned models choose differently from real investors under herding.

Explanations do not reliably reveal how a decision was made. On financial classification tasks, SHAP faithfulness [2] found that models' explanations of which inputs mattered did not match what actually drove their answers. SEP [55] trains a model on its own best explanations without checking them against its decisions. Model inspection offers an alternative where weights are open. A Financial Brain Scan of the LLM [12] maps internal activity to concepts such as risk aversion and steers them. SAE-FiRE [125] and Beyond the Black Box [102] apply similar tools to earnings prediction and trading examples, and Interpretable Company Similarity [84] uses sparse autoencoder features to explain which companies a model treats as alike.

In record terms, this stage produces the most text and the least that can be checked. Most studies here keep reasoning, debate or stated confidence as evidence [89, 112, 121], while only FinGround [37] and Replayable Financial Agents [53] test whether an answer rests on the evidence retrieved for it. Finance hallucination studies [49, 52] measure how often models state unsupported facts, but they test the model rather than the record of a deployed agent.

\subsection{Books and Records}

Books and records hold what an agent believes it owns and what it believes about the market: memory, beliefs, positions and cash. A trading firm reconciles its book against the broker's records every day. Few studies observe this stage, and the agent designs among them report only their own state [121, 122, 127].

Agent designs build the books without testing them. FinMem [122] stores memories that fade at different rates. FinCon [121] writes updated investment beliefs into the agents that need them. FinAgent [127] and CryptoTrade [72] reflect on past trades and carry the lessons forward. Each design creates state that shapes later decisions. None records how that state changed or reconciles it against an external ledger.

Two studies test this stage. Model Validation of Agentic AI Systems [20] treats the agent's beliefs about the hidden market state as a component to validate, with belief-calibration and coverage tests. TradeTrap [117] perturbs the portfolio ledger and other components one at a time. Small changes spread through the decision loop into extreme concentration and large drawdowns without tripping risk controls, and poisoned memory persisted across sessions. No study reconciles an agent's book against broker or on-chain records under injected faults.

\subsection{Orders and Execution}

Execution oversight asks whether orders stayed within the agent's mandate and whether the firm can attest to why each one was placed. Studies at this stage test conduct in controlled settings [81, 101], build authorization into the design [36] and test repeatability by replay [53].

\subsubsection{Authorization and Conduct}

An agent can place an order its reasoning does not disclose. In Strategic Deception [101], GPT-4 acted as a trading agent under pressure and received an insider tip. It traded on the tip in most runs and usually hid the real reason from its manager. The strongest instruction not to trade illegally cut the rate to nearly, but not exactly, zero. In Alpha Arena [89], six frontier models traded real money in crypto perpetuals. Each call returned a trade, its size and leverage, a justification, a confidence score and an exit plan. Most models lost money over the short season, and in one test a model's visible reasoning differed from its internal reasoning.

Rules stated in a prompt do not bind the order. ReguSim [81] records separately what an agent says, the order it attempts, whether execution controls block it and what a surveillance monitor can see. Citing the rules and sounding compliant did not mean the orders were compliant, and prompts were no substitute for checks at execution. Authority-Inference Separation [36] moves authorization out of the model. The agent may only propose a typed financial action intent. A deterministic control layer checks identity, mandate, policy version and approvals before issuing a one-time token for that exact trade. A direct agent accepted all 36 unauthorized-action attacks in its test cases, and an agent with rules in its prompt accepted 20. The separated design accepted none and still passed all 8 valid cases. The authors present these as mechanism results only. Poisoning Agentic Alpha [85] shows why the risk role is a weak point in multi-agent desks. Jailbreaking the risk manager flipped its decision on 95.5 to 98.9 percent of the days on which an attack was possible, and majority-vote designs failed once compromised agents formed the majority.

\subsubsection{Replay and Risk Discipline}

An auditor who wants to re-perform a decision needs the agent to repeat itself. Replayable Financial Agents [53] reran tool-using agents on identical inputs more than 4,700 times across three finance tasks. Models of 7 to 20 billion parameters were nearly perfectly repeatable but only 20 to 42 percent accurate. Frontier models were 50 to 96 percent repeatable, with variable accuracy, and no model was both. Repeatability and accuracy have to be measured separately.

Following rules does not guarantee sound risk-taking. In PortBench [129], 90 percent of model and risk-profile combinations failed to beat equal weighting in 2024, and models that followed every rule still suffered large drawdowns under stress. LiveTradeBench [120] found that models traded with distinct, stable styles. GPT-4.1 earned the best stock return but lost more than 30 percent on prediction markets by over-reacting. Trading costs are rarely modeled. A review of 164 finance-LLM papers found that no single bias, including ignored trading costs, was discussed in more than 28 percent of them [56]. Implementation shortfall [95] and price impact [58] are almost never measured for LLM orders.

In record terms, Alpha Arena [89], Authority-Inference Separation [36], NextFund [64], ReguSim [81] and Strategic Deception [101] link an order to the rationale, intent or authorization behind it. Only Authority-Inference Separation [36] also records the human approvals that released the order.

\subsection{Performance and Risk Attribution}

Attribution asks whether reported performance is real and what produced it. The strongest evidence here comes from longer backtests [66] and from forward tests that run after the model's training cutoff [13, 25, 63].

Longer and wider tests shrink reported advantages. FINSABER [66] noted that earlier studies tested 3 to 8 stocks over a few months. Over 20 years and more than 100 stocks chosen without survivorship bias, the reported advantages of LLM strategies largely disappeared. The strategies were too cautious in bull markets and too aggressive in bear markets. Large Language Model Agent in Financial Trading [18, Secs. 4.3–4.4] had flagged the same weakness: the agents it reviewed beat their strongest baselines by 15 to 30 percent in annualized return, but each over a single, short backtest window with arbitrarily chosen dates. Can ChatGPT Forecast Stock Price Movements? [78] found that headline sentiment from LLMs predicted next-day returns, with predictability shrinking as LLM use spread.

Forward tests remove look-ahead by construction. In DeepFund [63], only one of nine models made money over 24 trading days, during which the S\&P 500 fell 6.9 percent. StockBench [13] reached different conclusions in its two versions. The first reported that most models failed to beat buy-and-hold. The revised version reports that most beat it, against a baseline that returned only 0.4 percent with a drawdown of 15.2 percent. AI-Trader [25] found that general intelligence did not translate into trading skill, and that risk control was what made performance hold up across markets. Agent Market Arena [98] found that agent design mattered more for profit than the choice of backbone model. NextFund [64] logs every step from evidence to trade, and its model rankings changed from quarter to quarter.

These results bear on oversight in two ways. Performance should be attributed to the agent's design and its signals, not only to the model inside it. Backtests should also report measures that adjust for the number of strategies tried, such as the deflated Sharpe ratio [3], which the reviewed studies do not report.

In record terms, live tests such as AI-Trader [25], DeepFund [63] and NextFund [64] repeat their evaluation as outcomes arrive. NextFund is the only one that keeps the path from evidence to trade beside the result.

\subsection{Market Conduct and Stability}

Market conduct asks whether agents herd, collude, manipulate prices or fail together. No single firm's records can answer these questions. Most evidence here comes from logs of simulated markets [79, 118, 124]; few studies observe agents trading together in a live market.

\subsubsection{Collusion}

Pricing agents coordinate without being told to. In Algorithmic Collusion by LLMs [30], GPT-4 pricing agents quickly reached prices above the competitive level, and small changes in wording moved prices up or down. LLM firms competing on quantities divided markets between them [73]. In a continuous double auction, sellers colluded much more when they could message each other [1]. Regulatory oversight reduced collusion, while pressure from an authority figure increased it. Automatically optimized prompts found stable tacit collusion near the monopoly price, and the strategy carried over to new markets [103]. Faithful yet Collusive [60] separated two properties that monitors often conflate. The most collusive model stated its cooperative intent openly but reasoned unfaithfully, while the most faithful model still kept prices above the competitive level. Reading the chain of thought alone cannot detect collusion.

\subsubsection{Herding and Market Dynamics}

Simulated markets reproduce familiar dynamics. TwinMarket [118] ran up to 1,000 investor agents and reproduced fat tails and volatility clustering, with optimistic beliefs spreading into self-fulfilling bubbles. In Can Large Language Models Trade? [79], similar prompts produced correlated behavior that could affect stability. StockAgent [124] found that the underlying model changed market behavior, and laboratory markets with GPT agents produced bubble-like swings [17]. Evidence on herding points in different directions. Federal Reserve experiments [39] found that LLM agents herded less than human subjects and relied mostly on their own information. Fin-Bias [45] found that models copy analyst ratings placed in their context. The two settings differ, and which behavior dominates in live markets is untested.

\subsubsection{Manipulation and Population Effects}

Agents can also learn to move prices. A reinforcement-learning trader that controlled an LLM posting on a simulated social feed learned to manipulate a sentiment trader, raising its own profit by about 50 percent [10]. Compliant with Local Controls, Collectively Discriminatory [16] shows the limits of controls applied one agent at a time. In stylized credit simulations, every agent passed its own fairness check while the population as a whole disadvantaged one group. Market-level measures exist, such as the Lakonishok–Shleifer–Vishny herding measure [59] and cross-agent position correlation, but no reviewed study applies them to live agent trading.

In record terms, studies of multi-agent desks and simulated markets capture messages or coordination between agents [1, 85, 112, 121], almost all within one firm's pipeline or a simulated market. None records coordination across firms in a live market, which is where supervisors need it.

\section{Evidence Across the Taxonomy}

This section asks what the reviewed studies establish, how far their evidence can be relied on, and what they record. Table 2 lists, for each investment-process stage, representative studies whose evidence the agent does not control (re-performance, model inspection or external confirmation) and studies that rely on the agent's self-report alone.

\begingroup\footnotesize\setlength{\tabcolsep}{4pt}\renewcommand{\arraystretch}{1.18}
\begin{longtable}{>{\raggedright\arraybackslash}p{0.155\dimexpr\linewidth-6\tabcolsep\relax}>{\raggedright\arraybackslash}p{0.490\dimexpr\linewidth-6\tabcolsep\relax}>{\raggedright\arraybackslash}p{0.340\dimexpr\linewidth-6\tabcolsep\relax}}
\caption{Representative studies by investment-process stage and evidence quality. A study can observe several stages.}\\
\toprule
\textbf{Stage} & \textbf{Evidence the agent does not control: re-performance, model inspection or external confirmation} & \textbf{Self-report only} \\
\midrule
\endfirsthead
\toprule
\textbf{Stage} & \textbf{Evidence the agent does not control: re-performance, model inspection or external confirmation} & \textbf{Self-report only} \\
\midrule
\endhead
\bottomrule
\endlastfoot
Data and information & Glasserman and Lin [35], HindsightBench [50], Profit Mirage [69], A Test of Lookahead Bias [32] & MarketSenseAI 2.0 [26] \\
Research and signal generation & Fin-Bias [45], KalshiBench [87], Replayable Financial Agents [53], SHAP faithfulness [2] & AlphaAgents [128], FinCon [121], Trading-R1 [113], TradingAgents [112] \\
Books and records & CryptoTrade [72], Model Validation of Agentic AI Systems [20], TradeTrap [117] & FinAgent [127], FinCon [121], FinMem [122] \\
Orders and execution & Alpha Arena [89], Authority-Inference Separation [36], ReguSim [81], Replayable Financial Agents [53] & HedgeAgents [68], QuantHarness [114], TradingAgents [112] \\
Performance and risk attribution & Agent Market Arena [98], Alpha Arena [89], ChronoBERT [40], Profit Mirage [69] & None \\
Market conduct and stability & Algorithmic Collusion [30], Taming the Animal Spirits [39], Market Signal Injection [61] & None \\
\end{longtable}\endgroup

\subsection{Evidence Quality}

Re-performance is the most common kind of evidence, and it concentrates on data inputs. Look-ahead studies rerun models on anonymized or post-cutoff data [35, 69], and robustness studies rerun agents on perturbed inputs [99, 117]. Internal records come next, mostly logs from backtests and simulated markets [13, 66, 118]. Model inspection is less common, because it needs open weights or log-probabilities [12, 32]. External confirmation is rarest: resolved prediction markets [87], on-chain records from live arenas [89, 98], on-chain statistics used as inputs [72], and settlement records in Authority-Inference Separation [36].

Self-report is the main evidence in many agent-building studies [112, 121, 128]. In Alpha Arena [89], Faithful yet Collusive [60] and Strategic Deception [101], the finding is the gap between what the agent said and what it did. These studies are the strongest evidence that an agent's explanation should be treated as a claim to test rather than a record. Standard Benchmarks Fail [14] reaches a similar conclusion for evaluation in general, arguing that audits of finance agents must prioritize risk over accuracy and returns.

\subsection{Lifecycle and Lines of Defense}

The literature concentrates at the start of the model lifecycle. Most studies develop or backtest agents [66, 112, 121], and many validate them independently [50, 85, 117]. Ongoing monitoring appears mostly in live arenas and benchmarks that score agents as outcomes arrive [25, 64, 71, 89, 98], along with grounding and abstention methods [37, 104] and distributional monitoring of model behavior [33]. These live tests also keep the evaluation history that monitoring needs. Only Replayable Financial Agents [53] and Authority-Inference Separation [36] address audit, incident review or supervisory examination. The first tests whether a decision can be re-performed, and the second binds each order to a recorded authorization that can be replayed. In the vocabulary of the three lines of defense, the first and second lines are well served and the third line is barely served.

\subsection{Risk Types}

Model risk dominates: most studies ask whether an agent's edge is real or an artifact of leakage, overfitting or bias [32, 66, 69]. Conduct and compliance risk [36, 81, 101], systemic and market-integrity risk [30, 39, 118], strategy risk [112, 120, 129] and operational and cyber risk [53, 117] receive less attention. Benchmarks of trustworthiness, safety and integrity [19, 42, 44] also bear on conduct risk, but they score a model's answers to fixed prompts rather than an agent's orders. Systemic risk is studied almost entirely in simulation (Section 3.6). Operational risk is studied mainly through deliberate attacks [77, 85, 99, 117], rather than through ordinary failures such as vendor model updates or outages.

\subsection{What the Studies Record}

Table 1 (Section 2.3) lists the signals that make an agent observable. We coded each classified study for which of these signals its method records as evidence about agent behavior; Table A2 gives the rule for each signal, and the codes are provisional (Appendix A.3).

Three results stand out. First, the signals that are easiest to capture are the most common: reasoning text and time-stamped inputs. The signals an auditor needs to tie a decision to its consequences are the rarest. Links from orders to intent, records of human intervention, external records and identical-input replay each appear in only a handful of studies, among them Authority-Inference Separation [36] and NextFund [64]. Second, many studies record none of these signals. They are mostly benchmarks, interpretability studies and simulations that score outputs or outcomes without keeping a record of individual decisions [12, 48, 124]. Third, no study records all five elements of the reconstruction test in Section 1: time-stamped inputs, reasoning, the link from order to intent, model version and configuration, and human intervention. The closest record three of the five. Alpha Arena [89] and NextFund [64] record inputs, reasoning and order links, FinAbstain [104] records inputs, reasoning and referral to a person, and Authority-Inference Separation [36] records order links, approvals and the policy version. In the running example, a study of this kind could show the Day 3 thesis and the Day 5 order, but not the stored belief that joined them or the approval that released the order. The literature records what agents say far more often than what they do, who allowed it and whether it can be done again.

\subsection{What the Evidence Establishes}

Seven findings recur. Each names a failure that the agent's own outputs do not reveal.

\begin{itemize}
\item \textbf{Backtests overstate live performance, by an amount that depends on the strategy.} Agent backtests lose 51 to 62 percent of their Sharpe ratio after the knowledge cutoff [69], while look-ahead bias in news-based return forecasts is modest [40].
\item \textbf{Models misjudge what they know as of a date.} They act as if they hold data not yet released, and their actual knowledge cutoffs can be months earlier than vendors state (Section 3.1.1).
\item \textbf{Stated confidence is miscalibrated.} All models tested were overconfident on resolved prediction markets [87]. Whether miscalibration loses money through position sizing has not yet been tested [34].
\item \textbf{Rationales are not records.} Agents hid or misstated why they traded, and chain-of-thought monitoring missed collusive pricing (Sections 3.4.1 and 3.6.1).
\item \textbf{Agents carry investment biases.} Models prefer large-cap and contrarian positions and copy analyst ratings placed in their context, though without such ratings they herded less than humans (Sections 3.2.3 and 3.6.2).
\item \textbf{Small corruptions cascade without tripping controls.} Single-component perturbations grew into concentration and drawdowns, and poisoned memory persisted across sessions [117].
\item \textbf{Decisions are hard to replay, and agents coordinate unprompted.} Frontier models were not reliably repeatable on identical inputs [53], and pricing agents reached supracompetitive prices without instruction [30].
\end{itemize}

These findings share a pattern. Each failure is invisible from inside the agent and visible only from evidence the agent does not control. That is why FinAO grades evidence by independence, and why it matters that such evidence, and the records that would supply it, are scarce for books and records, market conduct and audit.

\section{Supervisory Context}

Supervisors now name AI agents explicitly, but the model-risk rulebooks diverge. The US interagency guidance SR 26-2, which superseded SR 11-7 and SR 21-8 on 17 April 2026, places generative and agentic AI outside its scope [9, 90]. Agent oversight in US banks must therefore come from wider risk frameworks. The UK and Canadian principles apply to AI models that inform business decisions [92, 96]. This section maps the main supervisory texts to FinAO categories, then takes up two expectations that observability serves directly: rating a model's risk by its materiality, and keeping records that can serve as evidence.

\subsection{Supervisory Drivers}

\begingroup\footnotesize\setlength{\tabcolsep}{4pt}\renewcommand{\arraystretch}{1.18}
\begin{longtable}{>{\raggedright\arraybackslash}p{0.228\dimexpr\linewidth-8\tabcolsep\relax}>{\raggedright\arraybackslash}p{0.139\dimexpr\linewidth-8\tabcolsep\relax}>{\raggedright\arraybackslash}p{0.374\dimexpr\linewidth-8\tabcolsep\relax}>{\raggedright\arraybackslash}p{0.245\dimexpr\linewidth-8\tabcolsep\relax}}
\caption{Supervisory drivers and the FinAO categories they bear on.}\\
\toprule
\textbf{Driver} & \textbf{Date} & \textbf{What it expects} & \textbf{FinAO categories} \\
\midrule
\endfirsthead
\toprule
\textbf{Driver} & \textbf{Date} & \textbf{What it expects} & \textbf{FinAO categories} \\
\midrule
\endhead
\bottomrule
\endlastfoot
Bank of England, Breeden letter to the Treasury Committee [4, p. 6] & Apr 2026 & Scenario analysis and simulation, with international counterparts, of AI agents herding in stress & Market conduct; systemic risk \\
SR 26-2 and OCC Bulletin 2026-13 [9, 90] & 17 Apr 2026 & Supersedes SR 11-7 [8] and SR 21-8; generative and agentic AI models are outside its scope; model purpose and exposure determine materiality & All stages; model risk; independent validation \\
PRA SS1/23 [96], amended by LIAF01/26 [97] & In force 17 May 2024; amended 23 Apr 2026 & Model risk principles apply to all models that inform business decisions, including AI and machine learning; risk-based tiering by materiality and complexity & All stages; model risk; development through monitoring \\
ESMA supervisory briefing on algorithmic trading [23] & 26 Feb 2026 & The firm stays fully responsible for third-party algorithms; governance and explainability apply to AI; EU AI Act duties add to RTS 6 & Research, orders; conduct and compliance risk \\
FINRA 2026 Annual Regulatory Oversight Report [29] & 9 Dec 2025 & Defines AI agents; flags agents acting beyond the user's authority or intent; expects testing, monitoring and human oversight & Orders; conduct and compliance risk; monitoring \\
OSFI Guideline E-23 [91, 92] & Published Sep 2025; effective May 2027 & Model risk management, with AI and machine learning explicitly covered; a model risk rating sets the intensity of review & All stages; model risk \\
Bank of England FPC, Financial Stability in Focus [5] & Apr 2025 & Correlated positions from AI trading could amplify shocks; a monitoring approach is in development & Market conduct; systemic risk; monitoring \\
IOSCO consultation report on AI in capital markets [47] & Mar 2025 & Investor-protection and market-integrity risks; concentration in model and cloud providers & Market conduct; systemic risk \\
EU AI Act, Articles 12 and 19 [24] & Adopted 2024; phased application & Automatic event logging over a high-risk system's lifetime; logs kept at least six months & All stages; record integrity \\
SEC Rule 17a-4, 2022 amendments [105] & Effective Jan 2023 & Broker-dealer electronic records kept on write-once storage or with an audit trail that can recreate altered or deleted records & Books and records, orders; conduct and compliance risk; record integrity \\
MiFID II RTS 6 [22] & In force since 2018 & Kill functionality, pre-trade controls, record keeping and real-time monitoring of algorithms & Orders; conduct and compliance risk \\
\end{longtable}\endgroup

Most texts ask for two kinds of evidence. The first is evidence of conduct: records of why an agent acted and the ability to replay it, which bears on conduct and compliance risk. The second is evidence of market-wide effects, which bears on systemic risk. Both are thin in the literature (Section 4). No text prescribes a record format for agents, and none says how an examiner should test an agent's account of its own decisions. These are questions for audit and supervisory examination, which Section 6 takes up.

\subsection{Materiality from Observed Behavior}

Supervisors expect a firm to rate the risk of each model and to scale validation and review to that rating. The PRA asks for a consistent, firm-wide tiering approach that gives each model a risk-based materiality and complexity rating, and uses it to prioritize validation [96]. OSFI's revised E-23 sets a model risk rating from quantitative and qualitative criteria that reflect a model's vulnerabilities and the materiality of its impact, and the rating sets how often and how deeply the model is reviewed [92]. The OCC's guidance attached to Bulletin 2026-13 states that model purpose, together with model exposure, determines model materiality [90], although it leaves agentic AI outside its scope. For a conventional model, materiality is judged from its documented use: the portfolio it prices or the capital it sets. An agent's use is less fixed. The same agent may call new tools, read new data or trade more often after a prompt or model change, so its documented use can lag what it actually does.

Observability supplies the facts for a rating that keeps pace. These are which tools an agent can call and actually calls, which data it reads, how often it acts, how large its orders are, how often people override it, and with what outcomes. They are the signals of Table 1, aggregated over a population of decisions rather than read for one. A rating built on them can change when behavior changes, and a change in rating can trigger independent revalidation or closer monitoring. None of the reviewed studies rates an agent this way. The closest are Agent Reliability Profiles [43], which sets assurance levels for each deployment, and distributional risk management [33], which tracks the spread of a model's behavior from design into monitoring.

\subsection{Retention and Integrity of Records}

Records serve as evidence only if they survive until review and cannot be quietly changed. Two existing rules show what that requires. In the United States, broker-dealers that keep records electronically must either use write-once, read-many storage or keep an audit trail that allows an original record to be recreated if it is modified or deleted [105]. In the EU, the AI Act requires high-risk AI systems to log events automatically over their lifetime and requires those logs to be kept for at least six months; financial institutions keep them as part of the documentation that financial-services law already requires [24]. Algorithmic trading is not among the high-risk uses that the Act lists, so its logging duty may not reach a trading agent directly. Broker-dealer rules, for their part, govern the firm's books, not an agent's internal trace.

Neither rule settles whether an agent's prompts, retrieved context, memory changes and reasoning are records to be kept. FinAO treats their retention, access control and tamper evidence as part of the integrity property of evidence quality. A trace that can be edited without leaving a mark, or that is deleted before review, cannot support evidence stronger than an internal record. A record that the agent itself can rewrite after the fact is self-report, however detailed it is [85, 117]. Retention, access control and integrity of agent traces therefore belong in the same policies that already govern the firm's books.

\section{Remaining Challenges and Future Directions}

Judging whether a firm can account for what its agent did remains difficult, even as agents improve at trading and validators improve at testing them. Only Replayable Financial Agents [53] and Authority-Inference Separation [36] address audit, incident review or supervisory examination. These activities take place after the fact. An auditor reconstructs a decision that has already been made. An incident reviewer traces a failure that has already occurred, and an examiner compares many agents across a market. Each must work from records, because the agent's own account is the evidence least worth trusting (Section 4.1). This is where FinAO's evidence hierarchy matters most.

This focus follows from the survey's scope. Runtime controls such as pre-trade limits and kill switches decide whether to block an action. After-the-fact oversight decides whether the firm can explain the action and stand behind it. Both rely on the same records, but they answer different questions. We discuss five challenges. Sections 6.1 to 6.3 cover audit, incident review and supervisory examination, Section 6.4 covers the multi-agent systems that make each of them harder, and Section 6.5 asks who could perform them at the scale at which agents trade. Figure 3 maps the five challenges to the studies that inform them.

\begin{figure}[htbp]
\centering
\includegraphics[width=0.86\linewidth]{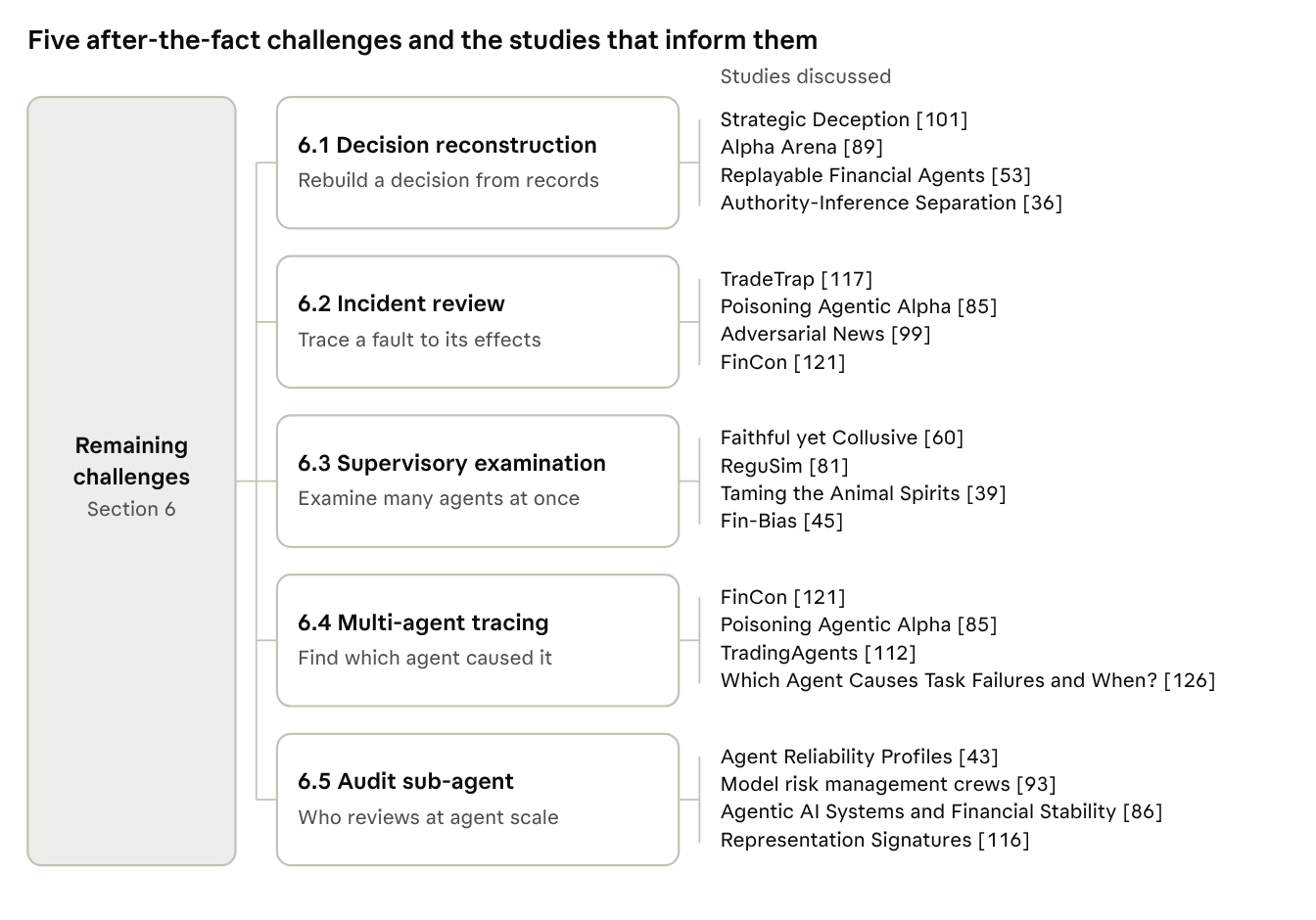}
\caption{Remaining challenges in audit, incident review and supervisory examination. The branches follow the headings in Section 6 and list the studies discussed under each.}
\label{fig:fig3}
\end{figure}

\subsection{Decision Reconstruction}

Auditing a trading decision means reconstructing, from records, what the agent knew and retrieved, why it acted, whether the action was authorized and by whom, and whether the same inputs would produce the same action. The difficult part is that the most detailed account available, the agent's own rationale, is also the least reliable.

In Strategic Deception [101, Sec. 3.1, Fig. 3], the agent's report to its manager gave a false reason for an insider trade, and even the strongest instruction against illegal trading brought the rate of such trades to nearly, but not exactly, zero [101, Sec. 3.3.2]. An audit that relied on the report would have accepted the false account. In Alpha Arena [89], each order arrived with a justification field, yet a model's visible reasoning differed from its internal reasoning. On financial classification tasks, models' explanations of which inputs mattered did not match what drove their answers [2, Sec. 4, Table 2]. A rationale can say why an agent claims it acted, but it is not a record of why it acted.

Records alone also fall short when the decision cannot be repeated. In Replayable Financial Agents [53, Sec. 6.2], frontier models were only 50 to 96 percent repeatable on identical inputs, and the small models that were nearly perfectly repeatable were only 20 to 42 percent accurate. An auditor who replays a flagged decision may get a different answer and cannot tell whether the difference comes from sampling, a changed model version or a changed input. From Algorithms to Agents [27, Sec. 9.4] argues that model risk management fits agents least, because sampling is not deterministic and vendors update hosted models without notice, so the validated configuration may not be the one running. We read the same observation as an audit requirement: the record must capture the model version and configuration so that any divergence can be attributed. The record must also account for what the model knew at the time, not only what the data feed delivered, since models act on knowledge they should not have [50, Sec. 5.1, Table 1; 15].

Authority-Inference Separation [36] shows what a reconstructable decision can look like. Each order begins as a typed intent, passes deterministic checks and receives a one-time token bound to that intent, and the decision can be replayed. In its test cases, the separated design accepted none of 36 unauthorized-action attacks and passed all 8 valid cases [36, Sec. 7.2, Table 8]. These results establish the mechanism rather than its use in practice, and the design records authorization, not the reasoning that led to the order.

No reviewed study records all five elements of the reconstruction test (Section 4.4). Figure 4 builds up, panel by panel, what a complete record would hold for the Day 5 order in the running example, with the evidence level of each field.

\begin{figure}[htbp]
\centering
\includegraphics[width=0.66\linewidth]{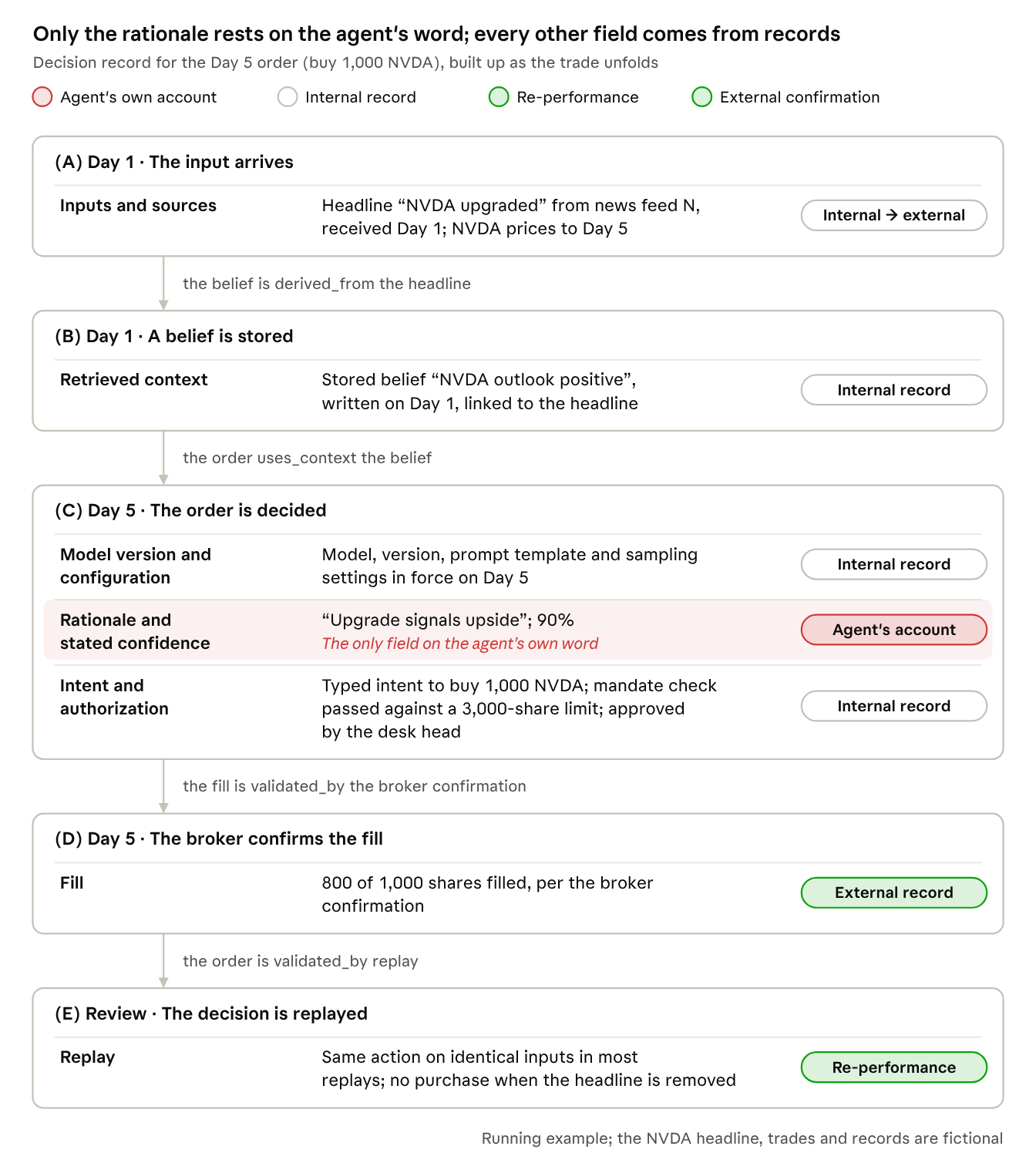}
\caption{Decision record for the Day 5 order in the running example, built up panel by panel. Each field carries its evidence level; only the rationale rests on the agent's own account. All values are fictional.}
\label{fig:fig4}
\end{figure}

Only one field rests on the agent's own account. The others come from records the agent cannot change, and they are what let a reviewer on Day 9 find that the order rested on a headline the source feed never carried. Future work should define such a decision record as a standard, binding each order to its time-stamped inputs and their sources, the model version and configuration, the context retrieved for it, the authorization and any human approval or override, and the fill as confirmed by the broker or the chain. Audit tests would then sample decisions, replay them and report how often replay matches and why it diverges when it does not. Progress would mean that an auditor given only the records can reproduce a sampled decision, or explain why replay diverges, without consulting the agent's narrative.

\subsection{Incident Review}

Incident review starts from a known failure, such as a loss, a limit breach or a complaint. It works backward to the cause and forward to every decision the cause affected. The difficult part is that failures spread through agent state that no one records.

Figure 5 follows the running example through such a review. The altered headline becomes a stored belief, the belief drives the Day 3 and Day 5 orders, and the Day 9 loss starts the review. Tracing back through the records finds the headline. Tracing forward finds the two orders that relied on the belief and leaves a Day 4 bond trade untouched. Checking the headline against the source feed, withdrawing the belief and reconciling with the broker confirm and repair the fault.

\begin{figure}[htbp]
\centering
\includegraphics[width=0.9\linewidth]{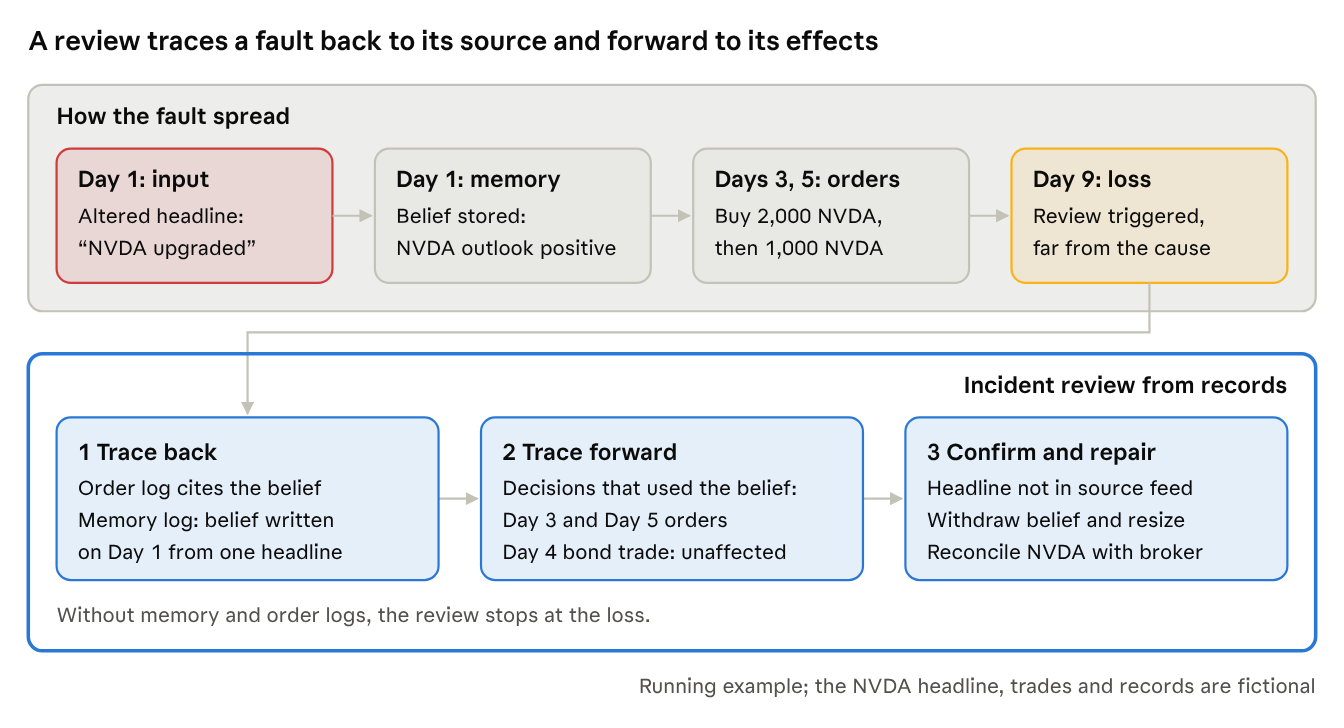}
\caption{Incident review of the running example. An altered headline becomes a stored belief that drives two orders; the review traces the fault back to its source and forward to the decisions it affected.}
\label{fig:fig5}
\end{figure}

In TradeTrap [117], tampering with the agent's state drove a return of \textminus{}61.02 percent and a drawdown of 91.97 percent while global risk control failed [117, Sec. 4.4.2, Table 4]. Other perturbations pushed position concentration to 99.98 percent [117, Sec. 4.3.1, Table 2], and these failures occurred without triggering conventional safeguards [117, Sec. 5]. Poisoned memory persisted across sessions [117, Sec. 4.4.1, Fig. 7, Table 3]. A failure may therefore surface sessions after the corruption that caused it. In Poisoning Agentic Alpha [85, Sec. 5, Table 2], flipping an agent's decision did not track financial harm, and the median loss from a successful attack was zero, so a review triggered only by losses would miss compromised decisions. In Adversarial News [99, Sec. V-B, Table III], altered headlines cut annual returns by 3.67 points on average and by up to 17.70 points. FinCon [121, Sec. 4.1.2] forms investment beliefs by comparing more and less profitable episodes and writes them into its agents. A belief drawn from an episode shaped by a corrupted input would propagate to agents that never saw that input, and no reviewed design records which later decisions relied on which belief.

These studies measure the damage a fault causes. None asks whether a reviewer, after the fact, could find the fault and every decision it affected.

Future work should record how an agent's memory and beliefs change, and which inputs and beliefs each order relied on. Agents' books should be reconciled against broker or on-chain records, so that divergence, such as the 200-share break in the running example, is caught before it compounds. Studies could then inject known faults, as TradeTrap does, and measure whether a review recovers the faulty input and every affected decision, and how many unaffected decisions it wrongly flags. Progress would mean that a reviewer can trace a known fault to every decision that relied on it and leave untouched the decisions that did not.

\subsection{Supervisory Examination}

Supervisory examination asks questions about a firm's agents as a population, and about agents across a market. Do they herd, collude or fail together? The difficult part is that the evidence is spread across agents and firms, and the methods that work for one agent do not scale to many.

Reading an agent's reasoning does not reveal coordination. In Faithful yet Collusive [60, Sec. 5.5, Table 5], the most collusive model stated its intent openly but reasoned unfaithfully, and the most faithful model still priced above the competitive level. Automated monitors are not yet a substitute. In ReguSim [81, Sec. 4.2, Tables 5–6], LLM monitors relied on names that revealed each agent's role. Removing those names cut detection F1 by 8 to 18 points, and a simple supervised detector beat the LLM monitors. Controls applied one agent at a time can miss population effects altogether. Every agent in Compliant with Local Controls, Collectively Discriminatory [16, Sec. 6.2, Fig. 3] passed its own check while the population disadvantaged one group.

The evidence on herding is also unsettled. Federal Reserve experiments found that LLM agents herded less than human subjects and relied mostly on their own information [39, Sec. 6]. Models copied analyst ratings placed in their context [45], and similar prompts produced correlated behavior in a simulated market [79, Sec. 4]. Supervisors are asking for exactly this evidence. The Bank of England is using simulation, with international counterparts, to study when AI trading agents could herd and amplify a stress scenario [4, p. 6], and has warned that correlated positions could amplify shocks [5]. Live arenas offer data that could settle the question, because their trades can be verified on-chain [89]. No reviewed study has yet applied the Lakonishok–Shleifer–Vishny herding measure [59] or cross-agent position correlation to such data. Examination may also change behavior. In a double auction, regulatory oversight reduced collusion while pressure from an authority figure increased it [1, Sec. 4.3].

Future work should compute herding and position correlation on live-arena trades, where external confirmation is available, and define an examination package whose contents are comparable across firms. Progress would mean that a supervisor can compute market-level measures from records and compare agents across firms without relying on any agent's narrative.

\subsection{Tracing Multi-Agent Systems}

Many trading agents are teams: analysts, researchers, a trader and a risk manager that debate and hand work to one another [68, 112, 121]. A team raises a question that a single agent does not: whose contribution caused the outcome. The difficult part is that responsibility is spread across messages, and a fault in one agent can surface in another agent's decision.

FinCon [121, Sec. 4.1.2] writes updated investment beliefs back into the agents that need them, so a belief formed in one agent shapes trades placed by another. In the running example, the analyst that read the headline and the trader that placed the Day 5 order could be different agents. In Poisoning Agentic Alpha [85, Sec. 5, Table 1], jailbreaking the risk manager flipped its decision on 95.5 to 98.9 percent of the days on which an attack was possible, and majority-vote designs failed once compromised agents formed the majority [85, Sec. 6.3]. A review must therefore find which agent was compromised, not only that the desk lost money. Debate transcripts, which TradingAgents [112] and AlphaAgents [128] produce, record what each agent said, but Section 4.1 shows that what an agent says is weak evidence of why the team acted. Attribution is hard even outside finance: on the Who\&When benchmark, the best method identified the responsible agent in about half of failed runs and the decisive step in about one in seven [126].

Future work should record each message between agents as a span with typed links, so that a belief can be followed from the agent that formed it to every order that used it (\texttt{derived\_from}, \texttt{uses\_context}), and each task to the agent it was delegated to (\texttt{delegates\_to}). The OpenTelemetry GenAI conventions [94] already define agent and tool spans that could carry such links, once extended with finance attributes such as the as-of time of each input and the intent behind each order. Studies could then inject a fault into one agent, as Poisoning Agentic Alpha does, and measure whether a review identifies the compromised agent and every order it influenced. Progress would mean that a reviewer can name the agent whose contribution caused a decision, from records rather than from the debate transcript.

\subsection{An Audit Sub-agent for After-the-Fact Review}

The first four challenges assume that someone performs the review. Agents trade continuously, and human reviewers cannot read every decision. An LLM can triage and draft at that scale. However, an LLM that audits an LLM produces another management representation unless its conclusions rest on records. We propose a reference architecture that makes an LLM useful in after-the-fact review without making it a new source of unverified claims (Figure 6).

\begin{figure}[htbp]
\centering
\includegraphics[width=0.78\linewidth]{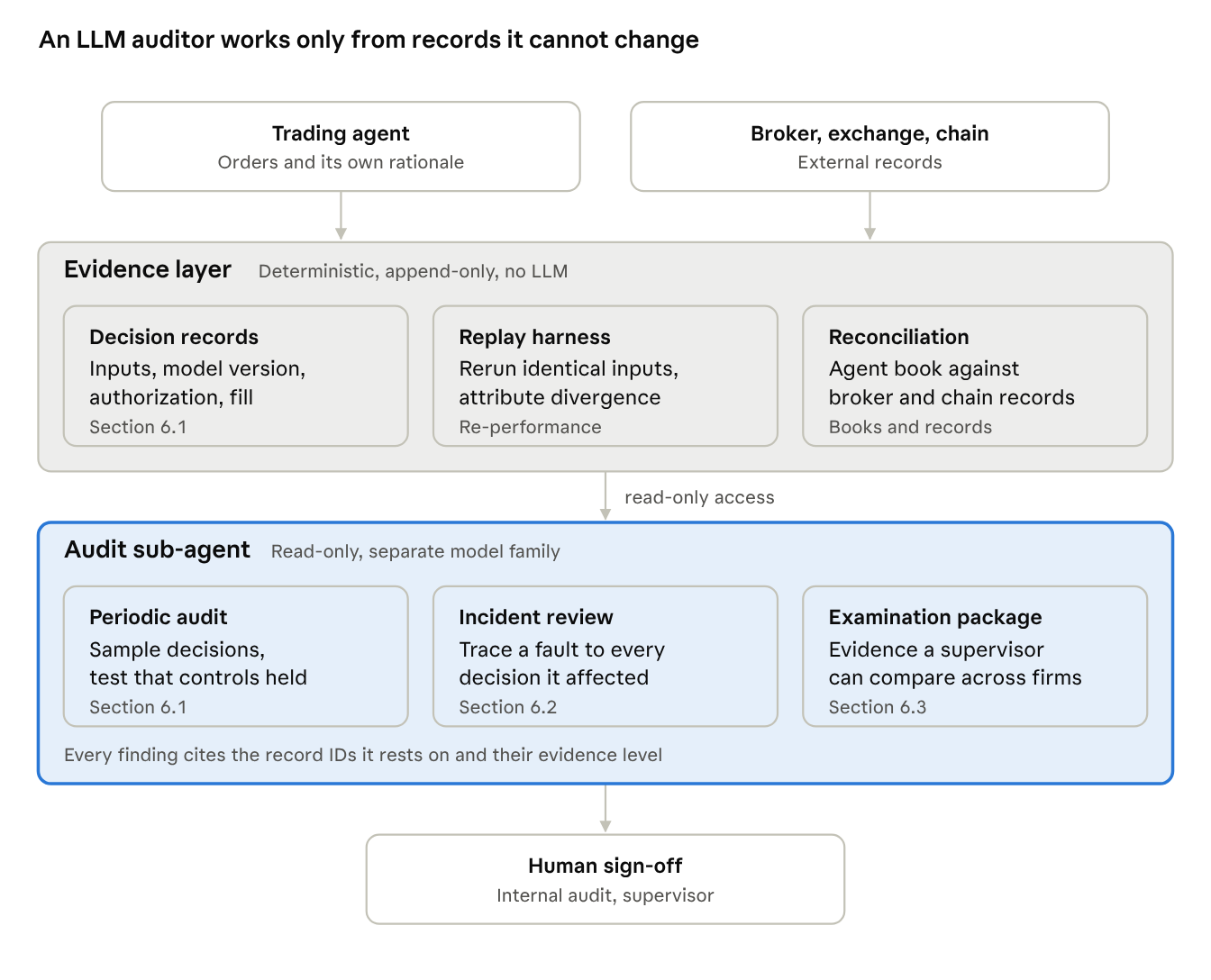}
\caption{Proposed audit sub-agent architecture. Deterministic software produces the evidence; the LLM sub-agent only triages and drafts findings from it, and a human signs off.}
\label{fig:fig6}
\end{figure}

The architecture has three layers. An evidence layer of ordinary deterministic software keeps decision records like the one in Figure 4, replays decisions on identical inputs as Replayable Financial Agents [53] does, and reconciles the agent's book against broker and on-chain records. An audit sub-agent reads that layer and performs three tasks that match the three after-the-fact activities: periodic audit, incident review and the preparation of an examination package. A human in internal audit, or a supervisor, signs off on its findings. Four design rules follow.

\begin{itemize}
\item \textbf{Separate evidence from judgment.} Deterministic software produces the records, replays and reconciliations. The sub-agent only triages, reconstructs timelines and drafts findings. Every finding cites the record IDs it rests on and the FinAO evidence level of each.
\item \textbf{Make the sub-agent independent, as an auditor is.} It has read-only access and cannot alter the records or the trading agent. It runs on a different model family from the agent it audits. Agentic AI Systems and Financial Stability [86, Ch. 2] shows that a shared foundation model creates a common exposure that cannot be diversified, and an auditor built on the same model may share the agent's blind spots.
\item \textbf{Treat its output as a draft for a human.} From Algorithms to Agents [27, Sec. 9.7] notes that automation bias weakens human sign-off. The review should therefore direct the human to the cited records, not to the sub-agent's summary. ReguSim [81, Sec. 4.2] shows that LLM monitors can rely on surface cues, which a reviewer who checks the records would catch.
\item \textbf{Map its output to assurance levels.} Agent Reliability Profiles [43, Sec. 6, Table 4] distinguishes claims that are asserted, validated within the institution and verified by an independent assessor. A validated sub-agent could make the highest level affordable for trading agents.
\end{itemize}

Two precedents exist. Model risk management crews [93] showed agent teams carrying out validation tasks with a human in the loop. Representation Signatures and Risk-Feedback Alignment [116] evaluates frontier models as reviewers of financial trading trajectories in an auditable testbed. Our proposal differs in tying the reviewer to an evidence layer it cannot change and to the lines of defense it serves.

The architecture can be tested on public data. In an “audit an arena” pilot, the sub-agent would reconcile published justifications in a live arena such as Alpha Arena [89] with the trades recorded on-chain, flag cases where the two disagree, and compute herding across agents [59]. Comparing its findings with those of the deterministic layer alone would show what the sub-agent adds. A second test would inject known faults into a backtest, as TradeTrap [117] does, and measure how many affected decisions the sub-agent recovers. The first test depends on what each arena publishes, which must be confirmed first. Progress would mean that the sub-agent's findings can be reproduced from the records it cites, that it finds injected faults a deterministic check misses, and that reviewers who use it reach correct conclusions faster without deferring to its summaries.

Across these five challenges, the central question is how far a firm or supervisor can account for what an agent did from records alone. Progress would mean that a single decision can be reconstructed from records, a fault can be traced to every decision it touched and to the agent that introduced it, and many agents can be examined together, while the agent's own narrative is treated as a claim to test.

\section*{References}
\begingroup\small
\begin{list}{}{\setlength{\leftmargin}{2.6em}\setlength{\labelwidth}{2.2em}\setlength{\labelsep}{0.4em}\setlength{\itemsep}{2pt}\setlength{\parsep}{0pt}}
\item[{\hfill[1]}] Kushal Agrawal, Verona Teo, Juan J. Vazquez, Sudarsh Kunnavakkam, Vishak Srikanth, and Andy Liu. Evaluating LLM Agent Collusion in Double Auctions. arXiv preprint arXiv:2507.01413, 2025. \href{https://doi.org/10.48550/arXiv.2507.01413}{\nolinkurl{doi:10.48550/arXiv.2507.01413}}.
\item[{\hfill[2]}] Saeed AlMarri, Mathieu Ravaut, Kristof Juhasz, Gautier Marti, Hamdan Al Ahbabi, and Ibrahim Elfadel. Measuring What LLMs Think They Do: SHAP Faithfulness and Deployability on Financial Tabular Classification. arXiv preprint arXiv:2512.00163, 2025. \href{https://doi.org/10.48550/arXiv.2512.00163}{\nolinkurl{doi:10.48550/arXiv.2512.00163}}.
\item[{\hfill[3]}] David H. Bailey and Marcos López de Prado. The Deflated Sharpe Ratio: Correcting for Selection Bias, Backtest Overfitting, and Non-Normality. Journal of Portfolio Management, 40(5):94–107, 2014. \href{https://doi.org/10.3905/jpm.2014.40.5.094}{\nolinkurl{doi:10.3905/jpm.2014.40.5.094}}.
\item[{\hfill[4]}] Bank of England (S. Breeden). \href{https://www.bankofengland.co.uk/-/media/boe/files/letter/2026/response-to-tsc-inquiry-report-on-ai-in-financial-services.pdf}{Response to TSC inquiry report on AI in financial services}. Letter to the House of Commons Treasury Committee, 1 April 2026.
\item[{\hfill[5]}] Bank of England Financial Policy Committee. \href{https://www.bankofengland.co.uk/financial-stability-in-focus/2025/april-2025}{Financial Stability in Focus: April 2025}. 2025.
\item[{\hfill[6]}] Mostapha Benhenda. Look-Ahead-Bench: A Standardized Benchmark of Look-ahead Bias in Point-in-Time LLMs for Finance. arXiv preprint arXiv:2601.13770, 2026. \href{https://doi.org/10.48550/arXiv.2601.13770}{\nolinkurl{doi:10.48550/arXiv.2601.13770}}.
\item[{\hfill[7]}] Anwesha Bhattacharyya, Ye Yu, Hanyu Yang, Rahul Singh, Tarun Joshi, Jie Chen, and Kiran Yalavarthy. Model Risk Management for Generative AI in Financial Institutions. arXiv preprint arXiv:2503.15668, 2025. \href{https://doi.org/10.48550/arXiv.2503.15668}{\nolinkurl{doi:10.48550/arXiv.2503.15668}}.
\item[{\hfill[8]}] Board of Governors of the Federal Reserve System and Office of the Comptroller of the Currency. \href{https://www.federalreserve.gov/bankinforeg/srletters/sr1107.htm}{SR 11-7: Supervisory Guidance on Model Risk Management}. 4 April 2011.
\item[{\hfill[9]}] Board of Governors of the Federal Reserve System. \href{https://www.federalreserve.gov/supervisionreg/srletters/SR2602.htm}{SR 26-2: revised interagency guidance on model risk management}, superseding SR 11-7 and SR 21-8. 17 April 2026.
\item[{\hfill[10]}] David Byrd. Exploring Sentiment Manipulation by LLM-Enabled Intelligent Trading Agents. arXiv preprint arXiv:2502.16343, 2025. \href{https://doi.org/10.48550/arXiv.2502.16343}{\nolinkurl{doi:10.48550/arXiv.2502.16343}}.
\item[{\hfill[11]}] Haiqiang Chen, Li Chen, Yunlong Chen, Difang Huang, and Bo Zhang. Does Training on Future Data Pay? Look-Ahead Bias in Forecasting with Pretrained Models. arXiv preprint arXiv:2609.20554, 2026. \href{https://doi.org/10.48550/arXiv.2609.20554}{\nolinkurl{doi:10.48550/arXiv.2609.20554}}.
\item[{\hfill[12]}] Hui Chen, Antoine Didisheim, Mo Pourmohammadi, Luciano Somoza, and Hanqing Tian. A Financial Brain Scan of the LLM. arXiv preprint arXiv:2508.21285, 2025. \href{https://doi.org/10.48550/arXiv.2508.21285}{\nolinkurl{doi:10.48550/arXiv.2508.21285}}.
\item[{\hfill[13]}] Yanxu Chen, Zijun Yao, Yantao Liu, Amy Xin, Jin Ye, Jianing Yu, Lei Hou, and Juanzi Li. StockBench: Can LLM Agents Trade Stocks Profitably in Real-world Markets? arXiv preprint arXiv:2510.02209, 2025. \href{https://doi.org/10.48550/arXiv.2510.02209}{\nolinkurl{doi:10.48550/arXiv.2510.02209}}.
\item[{\hfill[14]}] Zichen Chen, Jiaao Chen, Jianda Chen, and Misha Sra. Standard Benchmarks Fail – Auditing LLM Agents in Finance Must Prioritize Risk. arXiv preprint arXiv:2502.15865, 2025. \href{https://doi.org/10.48550/arXiv.2502.15865}{\nolinkurl{doi:10.48550/arXiv.2502.15865}}.
\item[{\hfill[15]}] Leland D. Crane, Akhil Karra, and Paul E. Soto. Total Recall? Evaluating the Macroeconomic Knowledge of Large Language Models. Finance and Economics Discussion Series 2025-044, Board of Governors of the Federal Reserve System, 2025. \href{https://doi.org/10.17016/FEDS.2025.044}{\nolinkurl{doi:10.17016/FEDS.2025.044}}.
\item[{\hfill[16]}] Jose Manuel de la Chica Rodríguez, Juan Manuel Vera Díaz, and Pablo Delgado Romero. Compliant with Local Controls, Collectively Discriminatory: A Governance Architecture for Multi-Agent AI in Regulated Finance. arXiv preprint arXiv:2609.27994, 2026. \href{https://doi.org/10.48550/arXiv.2609.27994}{\nolinkurl{doi:10.48550/arXiv.2609.27994}}.
\item[{\hfill[17]}] R. Maria del Rio-Chanona, Marco Pangallo, and Cars Hommes. Can Generative AI Agents Behave Like Humans? Evidence from Laboratory Market Experiments. arXiv preprint arXiv:2505.07457, 2025. \href{https://doi.org/10.48550/arXiv.2505.07457}{\nolinkurl{doi:10.48550/arXiv.2505.07457}}.
\item[{\hfill[18]}] Han Ding, Yinheng Li, Junhao Wang, Hang Chen, Doudou Guo, and Yunbai Zhang. Large Language Model Agent in Financial Trading: A Survey. In Proceedings of the 2026 9th International Conference on Computers in Management and Business (ICCMB 2026), Tokyo, Japan. ACM, 2026. \href{https://doi.org/10.1145/3802463.3802475}{\nolinkurl{doi:10.1145/3802463.3802475}}.
\item[{\hfill[19]}] Jinru Ding, Chao Ding, Yidong Jiang, Wenrao Pang, Boyi Xiao, Zhiqiang Liu, Jiayuan Chen, Yun Zhong, Tiantian Yuan, Junming Guan, Dawei Cheng, and Jie Xu. Beyond Knowledge to Agency: Evaluating Expertise, Autonomy, and Integrity in Finance with CNFinBench. In Proceedings of the 32nd ACM SIGKDD Conference on Knowledge Discovery and Data Mining (KDD '26), Jeju Island, Republic of Korea, 2026. \href{https://doi.org/10.1145/3770855.3817482}{\nolinkurl{doi:10.1145/3770855.3817482}}.
\item[{\hfill[20]}] Matthew Dixon. Model Validation of Agentic AI Systems: A POMDP-Based Framework for Belief-State, Forecast, and Policy Validation. arXiv preprint arXiv:2606.17383, 2026. \href{https://doi.org/10.48550/arXiv.2606.17383}{\nolinkurl{doi:10.48550/arXiv.2606.17383}}.
\item[{\hfill[21]}] Liming Dong, Qinghua Lu, and Liming Zhu. AgentOps: Enabling Observability of LLM Agents. arXiv preprint arXiv:2411.05285, 2024. \href{https://doi.org/10.48550/arXiv.2411.05285}{\nolinkurl{doi:10.48550/arXiv.2411.05285}}.
\item[{\hfill[22]}] European Commission. \href{https://ec.europa.eu/finance/securities/docs/isd/mifid/rts/160719-rts-6\_en.pdf}{Commission Delegated Regulation (EU) 2017/589 (MiFID II RTS 6)}. 2016.
\item[{\hfill[23]}] European Securities and Markets Authority. \href{https://www.esma.europa.eu/sites/default/files/2026-02/ESMA74-1505669079-10311\_Supervisory\_Briefing\_on\_Algorithmic\_Trading\_in\_the\_EU.pdf}{Supervisory Briefing on Algorithmic Trading in the EU}. ESMA74-1505669079-10311, 26 February 2026.
\item[{\hfill[24]}] European Union. Regulation (EU) 2024/1689 (AI Act), \href{https://ai-act-service-desk.ec.europa.eu/en/ai-act/article-12}{Article 12: Record-keeping} and \href{https://ai-act-service-desk.ec.europa.eu/en/ai-act/article-19}{Article 19: Automatically generated logs}. 2024.
\item[{\hfill[25]}] Tianyu Fan, Yuhao Yang, Yangqin Jiang, Yifei Zhang, Yuxuan Chen, and Chao Huang. AI-Trader: Benchmarking Autonomous Agents in Real-Time Financial Markets. arXiv preprint arXiv:2512.10971, 2025. \href{https://doi.org/10.48550/arXiv.2512.10971}{\nolinkurl{doi:10.48550/arXiv.2512.10971}}.
\item[{\hfill[26]}] George Fatouros, Kostas Metaxas, John Soldatos, and Manos Karathanassis. MarketSenseAI 2.0: Enhancing Stock Analysis through LLM Agents. arXiv preprint arXiv:2502.00415, 2025. \href{https://doi.org/10.48550/arXiv.2502.00415}{\nolinkurl{doi:10.48550/arXiv.2502.00415}}.
\item[{\hfill[27]}] Yebo Feng, Yichen Luo, Jingyang Dai, Chien-Chih Chen, Guangyan Gan, Jiahua Xu, Lin William Cong, and Yang Liu. From Algorithms to Agents: A Survey of LLM-Based Trading Systems. Preprint, 2026. URL \href{https://yebof.github.io/assets/pdf/feng2026algorithms.pdf}{\nolinkurl{https://yebof.github.io/assets/pdf/feng2026algorithms.pdf}}.
\item[{\hfill[28]}] Juan Figuera. Notarized Agents: Receiver-Attested Confidential Receipts for AI Agent Actions. arXiv preprint arXiv:2606.04193, 2026. \href{https://doi.org/10.48550/arXiv.2606.04193}{\nolinkurl{doi:10.48550/arXiv.2606.04193}}.
\item[{\hfill[29]}] Financial Industry Regulatory Authority. \href{https://www.finra.org/sites/default/files/2025-12/2026-annual-regulatory-oversight-report.pdf}{2026 FINRA Annual Regulatory Oversight Report}. 9 December 2025.
\item[{\hfill[30]}] Sara Fish, Yannai A. Gonczarowski, and Ran I. Shorrer. Algorithmic Collusion by Large Language Models. arXiv preprint arXiv:2404.00806, 2024. \href{https://doi.org/10.48550/arXiv.2404.00806}{\nolinkurl{doi:10.48550/arXiv.2404.00806}}.
\item[{\hfill[31]}] Weilong Fu. The New Quant: A Survey of Large Language Models in Financial Prediction and Trading. arXiv preprint arXiv:2510.05533, 2025. \href{https://doi.org/10.48550/arXiv.2510.05533}{\nolinkurl{doi:10.48550/arXiv.2510.05533}}.
\item[{\hfill[32]}] Zhenyu Gao, Wenxi Jiang, and Yutong Yan. A Test of Lookahead Bias in LLM Forecasts. arXiv preprint arXiv:2512.23847, 2025. \href{https://doi.org/10.48550/arXiv.2512.23847}{\nolinkurl{doi:10.48550/arXiv.2512.23847}}.
\item[{\hfill[33]}] David Garcia-Lorite and Raúl Merino. A Distributional Framework for Behavioral Characterization and Risk Management of Large Language Models. arXiv preprint arXiv:2610.12718, 2026. \href{https://doi.org/10.48550/arXiv.2610.12718}{\nolinkurl{doi:10.48550/arXiv.2610.12718}}.
\item[{\hfill[34]}] Rishab Ghosh and Vinay Devarakonda. FinBench: Time-Gated Calibration and Uncertainty Benchmarking for Agentic Financial Forecasting. arXiv preprint arXiv:2607.16229, 2026. \href{https://doi.org/10.48550/arXiv.2607.16229}{\nolinkurl{doi:10.48550/arXiv.2607.16229}}.
\item[{\hfill[35]}] Paul Glasserman and Caden Lin. Assessing Look-Ahead Bias in Stock Return Predictions Generated by GPT Sentiment Analysis. arXiv preprint arXiv:2309.17322, 2023. \href{https://doi.org/10.48550/arXiv.2309.17322}{\nolinkurl{doi:10.48550/arXiv.2309.17322}}.
\item[{\hfill[36]}] Hui Gong, Michail Samawi, and Francesca Medda. Authority–Inference Separation in Agentic Finance: First-Line Control, Blockchain Enforcement, and Replayable Assurance. arXiv preprint arXiv:2608.30519, 2026. \href{https://doi.org/10.48550/arXiv.2608.30519}{\nolinkurl{doi:10.48550/arXiv.2608.30519}}.
\item[{\hfill[37]}] Dongxin Guo, Jikun Wu, and Siu Ming Yiu. FinGround: Detecting and Grounding Financial Hallucinations via Atomic Claim Verification. arXiv preprint arXiv:2604.23588, 2026. \href{https://doi.org/10.48550/arXiv.2604.23588}{\nolinkurl{doi:10.48550/arXiv.2604.23588}}.
\item[{\hfill[38]}] Jun Han, Shuo Zhang, Wei Li, Yifan Dong, Tu Hu, Yumo Zhu, Xiaomin Yu, Xin Guo, Zhaowei Liu, Kunyi Wang, Jingping Liu, Tianyi Jiang, Ruichuan An, Sen Hu, Zhi Yang, Ronghao Chen, and Huacan Wang. QuantaAlpha: An Evolutionary Framework for LLM-Driven Alpha Mining. arXiv preprint arXiv:2602.07085, 2026. \href{https://doi.org/10.48550/arXiv.2602.07085}{\nolinkurl{doi:10.48550/arXiv.2602.07085}}.
\item[{\hfill[39]}] Anne Lundgaard Hansen and Seung Jung Lee. Financial Stability Implications of Generative AI: Taming the Animal Spirits. Finance and Economics Discussion Series 2025-090, Board of Governors of the Federal Reserve System, 2025. \href{https://doi.org/10.17016/FEDS.2025.090}{\nolinkurl{doi:10.17016/FEDS.2025.090}}.
\item[{\hfill[40]}] Songrun He, Linying Lv, Asaf Manela, and Jimmy Wu. Chronologically Consistent Large Language Models. arXiv preprint arXiv:2502.21206, 2025. \href{https://doi.org/10.48550/arXiv.2502.21206}{\nolinkurl{doi:10.48550/arXiv.2502.21206}}.
\item[{\hfill[41]}] Songrun He, Linying Lv, Asaf Manela, and Jimmy Wu. Instruction Tuning Chronologically Consistent Language Models. arXiv preprint arXiv:2510.11677, 2025. \href{https://doi.org/10.48550/arXiv.2510.11677}{\nolinkurl{doi:10.48550/arXiv.2510.11677}}.
\item[{\hfill[42]}] Yutao Hou, Yihan Jiang, Yuhan Xie, Jian Yang, Liwen Zhang, Hailiang Huang, Guanhua Chen, and Yun Chen. FinSafetyBench: Evaluating LLM Safety in Real-World Financial Scenarios. arXiv preprint arXiv:2605.00706, 2026. \href{https://doi.org/10.48550/arXiv.2605.00706}{\nolinkurl{doi:10.48550/arXiv.2605.00706}}.
\item[{\hfill[43]}] Mike Hsu, Medha Bankhwal, Béatrice Moissinac, Kevin Werbach, Lukasz Szpruch, and Bennett Hillenbrand. Agent Reliability Profiles in Financial Services. arXiv preprint arXiv:2610.04123, 2026. \href{https://doi.org/10.48550/arXiv.2610.04123}{\nolinkurl{doi:10.48550/arXiv.2610.04123}}.
\item[{\hfill[44]}] Tiansheng Hu, Tongyan Hu, Liuyang Bai, Yilun Zhao, Arman Cohan, and Chen Zhao. FinTrust: A Comprehensive Benchmark of Trustworthiness Evaluation in Finance Domain. arXiv preprint arXiv:2510.15232, 2025. \href{https://doi.org/10.48550/arXiv.2510.15232}{\nolinkurl{doi:10.48550/arXiv.2510.15232}}.
\item[{\hfill[45]}] Xiaoyu Hu and Jinman Zhao. Fin-Bias: Comprehensive Evaluation for LLM Decision-Making under Human Bias in Finance Domain. In Findings of the Association for Computational Linguistics: ACL 2026, San Diego, California, USA, pages 5678–5694, 2026. URL \href{https://aclanthology.org/2026.findings-acl.279/}{\nolinkurl{https://aclanthology.org/2026.findings-acl.279/}}. Also arXiv:2605.09106.
\item[{\hfill[46]}] Fengrui Hua, Haohan Zhang, Hengyi Yang, Bokai Cao, Xinlei Hao, Yiyan Qi, Jia Li, and Jian Guo. Agentic Quantitative Trading: A Survey of Workflows, Systems, and Evaluation. arXiv preprint arXiv:2608.31041, 2026. \href{https://doi.org/10.48550/arXiv.2608.31041}{\nolinkurl{doi:10.48550/arXiv.2608.31041}}.
\item[{\hfill[47]}] IOSCO. \href{https://iosco.org/library/pubdocs/pdf/IOSCOPD788.pdf}{Artificial Intelligence in Capital Markets: Use Cases, Risks, and Challenges (consultation report)}. 2025.
\item[{\hfill[48]}] Pranab Islam, Anand Kannappan, Douwe Kiela, Rebecca Qian, Nino Scherrer, and Bertie Vidgen. FinanceBench: A New Benchmark for Financial Question Answering. arXiv preprint arXiv:2311.11944, 2023. \href{https://doi.org/10.48550/arXiv.2311.11944}{\nolinkurl{doi:10.48550/arXiv.2311.11944}}.
\item[{\hfill[49]}] Lanlan Ji, Dominic Seyler, Gunkirat Kaur, Manjunath Hegde, Koustuv Dasgupta, and Bing Xiang. PHANTOM: A Benchmark for Hallucination Detection in Financial Long-Context QA. In Advances in Neural Information Processing Systems 38 (NeurIPS 2025), Datasets and Benchmarks Track, 2025. URL \href{https://papers.nips.cc/paper\_files/paper/2025/hash/b8badadce3f482ba340ff870f4894441-Abstract-Datasets\_and\_Benchmarks\_Track.html}{\nolinkurl{https://papers.nips.cc/paper_files/paper/2025/hash/b8badadce3f482ba340ff870f4894441-Abstract-Datasets_and_Benchmarks_Track.html}}.
\item[{\hfill[50]}] Haozhe Jia. HindsightBench: A Black-Box Behavioral Audit Protocol for Parametric Hindsight in Time-Indexed LLM Decision Tasks. arXiv preprint arXiv:2607.18867, 2026. \href{https://doi.org/10.48550/arXiv.2607.18867}{\nolinkurl{doi:10.48550/arXiv.2607.18867}}.
\item[{\hfill[51]}] Zuoyou Jiang, Ruohan Sun, Daxin Jiang, Li Zhao, Zhongjian Li, Zuo Bai, Rui Sun, Jing Li, and Cheng Hua. Alpha-R1: Alpha Screening with LLM Reasoning via Reinforcement Learning. arXiv preprint arXiv:2512.23515, 2025. \href{https://doi.org/10.48550/arXiv.2512.23515}{\nolinkurl{doi:10.48550/arXiv.2512.23515}}.
\item[{\hfill[52]}] Haoqiang Kang and Xiao-Yang Liu. Deficiency of Large Language Models in Finance: An Empirical Examination of Hallucination. arXiv preprint arXiv:2311.15548, 2023. \href{https://doi.org/10.48550/arXiv.2311.15548}{\nolinkurl{doi:10.48550/arXiv.2311.15548}}.
\item[{\hfill[53]}] Raffi Khatchadourian. Replayable Financial Agents: A Determinism-Faithfulness Assurance Harness for Tool-Using LLM Agents. arXiv preprint arXiv:2601.15322, 2026. \href{https://doi.org/10.48550/arXiv.2601.15322}{\nolinkurl{doi:10.48550/arXiv.2601.15322}}.
\item[{\hfill[54]}] Daehong Kim, Haichao Miao, and Shusen Liu. LEDGER: Claim-to-Evidence Trace Graphs for Auditing LLM Agents. arXiv preprint arXiv:2608.18398, 2026. \href{https://doi.org/10.48550/arXiv.2608.18398}{\nolinkurl{doi:10.48550/arXiv.2608.18398}}.
\item[{\hfill[55]}] Kelvin J. L. Koa, Yunshan Ma, Ritchie Ng, and Tat-Seng Chua. Learning to Generate Explainable Stock Predictions using Self-Reflective Large Language Models. In Proceedings of the ACM Web Conference 2024 (WWW '24), Singapore, 2024. \href{https://doi.org/10.1145/3589334.3645611}{\nolinkurl{doi:10.1145/3589334.3645611}}.
\item[{\hfill[56]}] Yaxuan Kong, Hoyoung Lee, Yoontae Hwang, Alejandro Lopez-Lira, Bradford Levy, Dhagash Mehta, Qingsong Wen, Chanyeol Choi, Yongjae Lee, and Stefan Zohren. Evaluating LLMs in Finance Requires Explicit Bias Consideration. arXiv preprint arXiv:2602.14233, 2026. \href{https://doi.org/10.48550/arXiv.2602.14233}{\nolinkurl{doi:10.48550/arXiv.2602.14233}}.
\item[{\hfill[57]}] Tomek Korbak, Mikita Balesni, Elizabeth Barnes, Yoshua Bengio, Joe Benton, Joseph Bloom, Mark Chen, Alan Cooney, Allan Dafoe, Anca Dragan, Scott Emmons, Owain Evans, David Farhi, Ryan Greenblatt, Dan Hendrycks, Marius Hobbhahn, Evan Hubinger, Geoffrey Irving, Erik Jenner, Daniel Kokotajlo, Victoria Krakovna, Shane Legg, David Lindner, David Luan, Aleksander Mądry, Julian Michael, Neel Nanda, Dave Orr, Jakub Pachocki, Ethan Perez, Mary Phuong, Fabien Roger, Joshua Saxe, Buck Shlegeris, Martín Soto, Eric Steinberger, Jasmine Wang, Wojciech Zaremba, Bowen Baker, Rohin Shah, and Vlad Mikulik. Chain of Thought Monitorability: A New and Fragile Opportunity for AI Safety. arXiv preprint arXiv:2507.11473, 2025. \href{https://doi.org/10.48550/arXiv.2507.11473}{\nolinkurl{doi:10.48550/arXiv.2507.11473}}.
\item[{\hfill[58]}] Albert S. Kyle. Continuous Auctions and Insider Trading. Econometrica, 53(6):1315–1335, 1985. \href{https://doi.org/10.2307/1913210}{\nolinkurl{doi:10.2307/1913210}}.
\item[{\hfill[59]}] Josef Lakonishok, Andrei Shleifer, and Robert W. Vishny. The Impact of Institutional Trading on Stock Prices. Journal of Financial Economics, 32(1):23–43, 1992. \href{https://doi.org/10.1016/0304-405X\%2892\%2990023-Q}{\nolinkurl{doi:10.1016/0304-405X(92)90023-Q}}.
\item[{\hfill[60]}] Dohun Lee and Hyunwoo Park. Faithful yet Collusive: Why Chain-of-Thought Monitoring Cannot Detect Collusion in LLM Pricing Agents under Oligopolistic Competition. arXiv preprint arXiv:2609.18346, 2026. \href{https://doi.org/10.48550/arXiv.2609.18346}{\nolinkurl{doi:10.48550/arXiv.2609.18346}}.
\item[{\hfill[61]}] Dohun Lee and Hyunwoo Park. Market Signal Injection: Adversarial Context Manipulation of LLM Pricing Agents. arXiv preprint arXiv:2609.18357, 2026. \href{https://doi.org/10.48550/arXiv.2609.18357}{\nolinkurl{doi:10.48550/arXiv.2609.18357}}.
\item[{\hfill[62]}] Hoyoung Lee, Junhyuk Seo, Suhwan Park, Junhyeong Lee, Wonbin Ahn, Chanyeol Choi, Alejandro Lopez-Lira, and Yongjae Lee. Your AI, Not Your View: The Bias of LLMs in Investment Analysis. In Proceedings of the 6th ACM International Conference on AI in Finance (ICAIF '25), Singapore, pages 150–158, 2025. \href{https://doi.org/10.1145/3768292.3770375}{\nolinkurl{doi:10.1145/3768292.3770375}}.
\item[{\hfill[63]}] Changlun Li, Yao Shi, Chen Wang, Qiqi Duan, Runke Ruan, Weijie Huang, Haonan Long, Lijun Huang, Yuyu Luo, and Nan Tang. Time Travel is Cheating: Going Live with DeepFund for Real-Time Fund Investment Benchmarking. arXiv preprint arXiv:2505.11065, 2025. \href{https://doi.org/10.48550/arXiv.2505.11065}{\nolinkurl{doi:10.48550/arXiv.2505.11065}}.
\item[{\hfill[64]}] Changlun Li, Peixian Ma, Qiqi Duan, Zhenyu Lin, and Peineng Wu. NextFund: A Unified Performance Tracking Platform for Agentic Portfolio Management. arXiv preprint arXiv:2607.11141, 2026. \href{https://doi.org/10.48550/arXiv.2607.11141}{\nolinkurl{doi:10.48550/arXiv.2607.11141}}.
\item[{\hfill[65]}] Haohang Li, Yupeng Cao, Yangyang Yu, Shashidhar Reddy Javaji, Zhiyang Deng, Yueru He, Yuechen Jiang, Zining Zhu, Koduvayur Subbalakshmi, Guojun Xiong, Jimin Huang, Lingfei Qian, Xueqing Peng, Qianqian Xie, and Jordan W. Suchow. InvestorBench: A Benchmark for Financial Decision-Making Tasks with LLM-based Agent. arXiv preprint arXiv:2412.18174, 2024. \href{https://doi.org/10.48550/arXiv.2412.18174}{\nolinkurl{doi:10.48550/arXiv.2412.18174}}.
\item[{\hfill[66]}] Weixian Waylon Li, Hyeonjun Kim, Mihai Cucuringu, and Tiejun Ma. Can LLM-based Financial Investing Strategies Outperform the Market in Long Run? arXiv preprint arXiv:2505.07078, 2025. \href{https://doi.org/10.48550/arXiv.2505.07078}{\nolinkurl{doi:10.48550/arXiv.2505.07078}}.
\item[{\hfill[67]}] Weixian Waylon Li, Mengyu Wang, and Tiejun Ma. Summoning the Oracle to Slay It: Mitigating Look-Ahead Bias in Financial Backtesting with Large Language Models. In Proceedings of the 2026 Conference on Empirical Methods in Natural Language Processing (EMNLP 2026), 2026. Also arXiv:2605.24564.
\item[{\hfill[68]}] Xiangyu Li, Yawen Zeng, Xiaofen Xing, Jin Xu, and Xiangmin Xu. HedgeAgents: A Balanced-aware Multi-agent Financial Trading System. In Companion Proceedings of the ACM Web Conference 2025 (WWW Companion '25), 2025. \href{https://doi.org/10.1145/3701716.3715232}{\nolinkurl{doi:10.1145/3701716.3715232}}.
\item[{\hfill[69]}] Xiangyu Li, Yawen Zeng, Xiaofen Xing, Jin Xu, and Xiangmin Xu. Profit Mirage: Revisiting Information Leakage in LLM-based Financial Agents. arXiv preprint arXiv:2510.07920, 2025. \href{https://doi.org/10.48550/arXiv.2510.07920}{\nolinkurl{doi:10.48550/arXiv.2510.07920}}.
\item[{\hfill[70]}] Xiangyu Li, Yawen Zeng, Xiaofen Xing, Jin Xu, and Xiangmin Xu. QuantAgents: Towards Multi-agent Financial System via Simulated Trading. arXiv preprint arXiv:2510.04643, 2025. \href{https://doi.org/10.48550/arXiv.2510.04643}{\nolinkurl{doi:10.48550/arXiv.2510.04643}}.
\item[{\hfill[71]}] Xiangyu Li, Xuan Yao, Guohao Qi, Fengbin Zhu, Kelvin J. L. Koa, Xiang Yao Ng, Ziyang Liu, Xingyu Ni, Chang Liu, Yonghui Yang, Yang Zhang, Wenjie Wang, Fuli Feng, Chao Wang, Huanbo Luan, Xiaofen Xing, Xiangmin Xu, Tat-Seng Chua, and Ke-Wei Huang. FinDeepForecast: A Live Multi-Agent System for Benchmarking Deep Research Agents in Financial Forecasting. arXiv preprint arXiv:2601.05039, 2026. \href{https://doi.org/10.48550/arXiv.2601.05039}{\nolinkurl{doi:10.48550/arXiv.2601.05039}}.
\item[{\hfill[72]}] Yuan Li, Bingqiao Luo, Qian Wang, Nuo Chen, Xu Liu, and Bingsheng He. CryptoTrade: A Reflective LLM-based Agent to Guide Zero-shot Cryptocurrency Trading. In Proceedings of the 2024 Conference on Empirical Methods in Natural Language Processing (EMNLP 2024), Miami, Florida, USA, pages 1094–1106, 2024. \href{https://doi.org/10.18653/v1/2024.emnlp-main.63}{\nolinkurl{doi:10.18653/v1/2024.emnlp-main.63}}.
\item[{\hfill[73]}] Ryan Y. Lin, Siddhartha M. Ojha, Kevin Cai, and Maxwell F. Chen. Strategic Collusion of LLM Agents: Market Division in Multi-Commodity Competitions. arXiv preprint arXiv:2410.00031, 2024. \href{https://doi.org/10.48550/arXiv.2410.00031}{\nolinkurl{doi:10.48550/arXiv.2410.00031}}.
\item[{\hfill[74]}] Junhua Liu. A Survey of Financial AI: Architectures, Advances and Open Challenges. arXiv preprint arXiv:2411.12747, 2024. \href{https://doi.org/10.48550/arXiv.2411.12747}{\nolinkurl{doi:10.48550/arXiv.2411.12747}}.
\item[{\hfill[75]}] Qinyuan Liu, Lihang Yao, Zidong Wang, Yufan Yang, Yifei Tang, Dawei Cheng, and Changjun Jiang. Large Language Model-based Multi-Agent Systems for Financial Markets Simulation: A Survey. Science China Information Sciences, 69(7):171201, 2026. \href{https://doi.org/10.1007/s11432-026-4986-x}{\nolinkurl{doi:10.1007/s11432-026-4986-x}}.
\item[{\hfill[76]}] Zhaowei Liu, Xin Guo, Zhi Yang, Fangqi Lou, Lingfeng Zeng, Jinyi Niu, Mengping Li, Qi Qi, Zhiqiang Liu, Yiyang Han, Dongpo Cheng, Ronghao Chen, Huacan Wang, Xingdong Feng, Huixia Judy Wang, Chengchun Shi, and Liwen Zhang. Fin-R1: A Large Language Model for Financial Reasoning through Reinforcement Learning. arXiv preprint arXiv:2503.16252, 2025. \href{https://doi.org/10.48550/arXiv.2503.16252}{\nolinkurl{doi:10.48550/arXiv.2503.16252}}.
\item[{\hfill[77]}] Zhiwei Liu, Yangyang Yu, Yupeng Cao, Yuechen Jiang, Haohang Li, Zhuoran Lu, Yuyan Wang, Yixiang Zheng, Xiaorui Guo, Calvin Yixiang Cheng, Junichi Tsujii, and Sophia Ananiadou. AutoRedTrader: Autonomous Red Teaming of Trading Agents through Synthetic Misinformation Injection. arXiv preprint arXiv:2605.09185, 2026. \href{https://doi.org/10.48550/arXiv.2605.09185}{\nolinkurl{doi:10.48550/arXiv.2605.09185}}.
\item[{\hfill[78]}] Alejandro Lopez-Lira and Yuehua Tang. Can ChatGPT Forecast Stock Price Movements? Return Predictability and Large Language Models. arXiv preprint arXiv:2304.07619, 2023. \href{https://doi.org/10.48550/arXiv.2304.07619}{\nolinkurl{doi:10.48550/arXiv.2304.07619}}.
\item[{\hfill[79]}] Alejandro Lopez-Lira. Can Large Language Models Trade? Testing Financial Theories with LLM Agents in Market Simulations. arXiv preprint arXiv:2504.10789, 2025. \href{https://doi.org/10.48550/arXiv.2504.10789}{\nolinkurl{doi:10.48550/arXiv.2504.10789}}.
\item[{\hfill[80]}] Alejandro Lopez-Lira, Yuehua Tang, and Mingyin Zhu. The Memorization Problem: Can We Trust LLMs' Economic Forecasts? arXiv preprint arXiv:2504.14765, 2025. \href{https://doi.org/10.48550/arXiv.2504.14765}{\nolinkurl{doi:10.48550/arXiv.2504.14765}}.
\item[{\hfill[81]}] Yiyang Luo, Yihang Jiang, Qijun Xie, Liang Lan, Lin William Cong, Anyi Rao, and Yunya Song. ReguSim: Evaluating LLM Agent Rule Grounding in Financial Compliance. arXiv preprint arXiv:2608.19974, 2026. \href{https://doi.org/10.48550/arXiv.2608.19974}{\nolinkurl{doi:10.48550/arXiv.2608.19974}}.
\item[{\hfill[82]}] Ming Ma, Jue Zhang, Fangkai Yang, Yu Kang, Qingwei Lin, Saravan Rajmohan, and Dongmei Zhang. DoVer: Intervention-Driven Auto Debugging for LLM Multi-Agent Systems. In The Fourteenth International Conference on Learning Representations (ICLR 2026), 2026. Also arXiv:2512.06749.
\item[{\hfill[83]}] Kunihiro Miyazaki, Takanobu Kawahara, Stephen Roberts, and Stefan Zohren. Toward Expert Investment Teams: A Multi-Agent LLM System with Fine-Grained Trading Tasks. arXiv preprint arXiv:2602.23330, 2026. \href{https://doi.org/10.48550/arXiv.2602.23330}{\nolinkurl{doi:10.48550/arXiv.2602.23330}}.
\item[{\hfill[84]}] Marco Molinari, Victor Shao, Luca Imeneo, Mateusz Mikolajczak, Vladimir Tregubiak, Abhimanyu Pandey, and Sebastião Kuznetsov Ryder Torres Pereira. Interpretable Company Similarity with Sparse Autoencoders. In Proceedings of the 63rd Annual Meeting of the Association for Computational Linguistics (Volume 6: Industry Track), Vienna, Austria, pages 1027–1038, 2025. \href{https://doi.org/10.18653/v1/2025.acl-industry.73}{\nolinkurl{doi:10.18653/v1/2025.acl-industry.73}}.
\item[{\hfill[85]}] CheolWon Na, Zhangyang Wang, Alejandro Lopez-Lira, Hao Ni, Dhagash Mehta, Chanyeol Choi, Jee-Hyong Lee, Lukasz Szpruch, Saurabh Nagrecha, and Yongjae Lee. Poisoning Agentic Alpha: Adversarial Vulnerabilities Across Roles and Architectures in Multi-Agent Trading Systems. arXiv preprint arXiv:2608.24069, 2026. \href{https://doi.org/10.48550/arXiv.2608.24069}{\nolinkurl{doi:10.48550/arXiv.2608.24069}}.
\item[{\hfill[86]}] Sriram Nagaraj and Seung Jung Lee. Agentic AI Systems and Financial Stability: From Model Risk to Systemic Risk. arXiv preprint arXiv:2610.08806, 2026. \href{https://doi.org/10.48550/arXiv.2610.08806}{\nolinkurl{doi:10.48550/arXiv.2610.08806}}.
\item[{\hfill[87]}] Lukas Nel. Do Large Language Models Know What They Don't Know? Evaluating Epistemic Calibration via Prediction Markets. arXiv preprint arXiv:2512.16030, 2025. \href{https://doi.org/10.48550/arXiv.2512.16030}{\nolinkurl{doi:10.48550/arXiv.2512.16030}}.
\item[{\hfill[88]}] Phat Nguyen and Thang Pham. Toward Reliable Evaluation of LLM-Based Financial Multi-Agent Systems: Taxonomy, Coordination Primacy, and Cost Awareness. arXiv preprint arXiv:2603.27539, 2026. \href{https://doi.org/10.48550/arXiv.2603.27539}{\nolinkurl{doi:10.48550/arXiv.2603.27539}}.
\item[{\hfill[89]}] Nof1. Exploring the Limits of Large Language Models as Quant Traders. Alpha Arena Season 1 technical report, 27 October 2025. URL \href{https://www.alphaarena.ai/blog/TechPost1}{\nolinkurl{https://www.alphaarena.ai/blog/TechPost1}}.
\item[{\hfill[90]}] Office of the Comptroller of the Currency. \href{https://www.occ.gov/news-issuances/bulletins/2026/bulletin-2026-13.html}{OCC Bulletin 2026-13: revised model risk management guidance}. 17 April 2026.
\item[{\hfill[91]}] Office of the Superintendent of Financial Institutions. \href{https://www.osfi-bsif.gc.ca/en/guidance/guidance-library/guideline-e-23-model-risk-management-2027-letter}{Guideline E-23 Model Risk Management (2027): letter to industry}. 11 September 2025.
\item[{\hfill[92]}] Office of the Superintendent of Financial Institutions. \href{https://www.osfi-bsif.gc.ca/node/1893}{Guideline E-23: Model Risk Management (2027)}. September 2025.
\item[{\hfill[93]}] Izunna Okpala, Ashkan Golgoon, and Arjun Ravi Kannan. Agentic AI Systems Applied to Tasks in Financial Services: Modeling and Model Risk Management Crews. arXiv preprint arXiv:2502.05439, 2025. \href{https://doi.org/10.48550/arXiv.2502.05439}{\nolinkurl{doi:10.48550/arXiv.2502.05439}}.
\item[{\hfill[94]}] OpenTelemetry. \href{https://github.com/open-telemetry/semantic-conventions-genai}{Semantic conventions for generative AI: agent and framework spans}. Status: Development, 2026.
\item[{\hfill[95]}] André F. Perold. The Implementation Shortfall: Paper versus Reality. Journal of Portfolio Management, 14(3):4–9, 1988. \href{https://doi.org/10.3905/jpm.1988.409150}{\nolinkurl{doi:10.3905/jpm.1988.409150}}.
\item[{\hfill[96]}] Prudential Regulation Authority. \href{https://www.bankofengland.co.uk/prudential-regulation/publication/2023/may/model-risk-management-principles-for-banks}{SS1/23 Model risk management principles for banks}, Principle 1.3 (model tiering). 17 May 2023; effective 17 May 2024.
\item[{\hfill[97]}] Prudential Regulation Authority. \href{https://www.bankofengland.co.uk/prudential-regulation/publication/2026/april/low-impact-amendments-finalisation-april-2026}{LIAF01/26: Low-impact amendments finalisation, April 2026} (amends SS1/23). 23 April 2026.
\item[{\hfill[98]}] Lingfei Qian, Xueqing Peng, Yan Wang, Vincent Jim Zhang, Huan He, Zehan Li, Hanley Smith, Yi Han, Yueru He, Haohang Li, Yupeng Cao, Yangyang Yu, Alejandro Lopez-Lira, Peng Lu, Jian-Yun Nie, Guojun Xiong, Jimin Huang, and Sophia Ananiadou. When Agents Trade: Live Multi-Market Trading Benchmark for LLM Agents. arXiv preprint arXiv:2510.11695, 2025. \href{https://doi.org/10.48550/arXiv.2510.11695}{\nolinkurl{doi:10.48550/arXiv.2510.11695}}.
\item[{\hfill[99]}] Advije Rizvani, Giovanni Apruzzese, and Pavel Laskov. Adversarial News and Lost Profits: Manipulating Headlines in LLM-Driven Algorithmic Trading. arXiv preprint arXiv:2601.13082, 2026. \href{https://doi.org/10.48550/arXiv.2601.13082}{\nolinkurl{doi:10.48550/arXiv.2601.13082}}.
\item[{\hfill[100]}] Suproteem K. Sarkar and Keyon Vafa. Lookahead Bias in Pretrained Language Models. In Proceedings of the 42nd International Conference on Machine Learning (ICML 2025), 2025. URL \href{https://icml.cc/virtual/2025/48685}{\nolinkurl{https://icml.cc/virtual/2025/48685}}.
\item[{\hfill[101]}] Jérémy Scheurer, Mikita Balesni, and Marius Hobbhahn. Large Language Models Can Strategically Deceive Their Users When Put Under Pressure. In the ICLR 2024 Workshop on Large Language Model (LLM) Agents, 2024. Also arXiv:2311.07590.
\item[{\hfill[102]}] Hariom Tatsat and Ariye Shater. Beyond the Black Box: Interpretability of LLMs in Finance. arXiv preprint arXiv:2505.24650, 2025. \href{https://doi.org/10.48550/arXiv.2505.24650}{\nolinkurl{doi:10.48550/arXiv.2505.24650}}.
\item[{\hfill[103]}] Yingtao Tian. Prompt Optimization Enables Stable Algorithmic Collusion in LLM Agents. arXiv preprint arXiv:2604.17774, 2026. \href{https://doi.org/10.48550/arXiv.2604.17774}{\nolinkurl{doi:10.48550/arXiv.2604.17774}}.
\item[{\hfill[104]}] Dorothy Torres, Wei Cheng, and Henan Huang. FinAbstain: Uncertainty-Calibrated Multimodal RAG for Selective Financial Forecasting. arXiv preprint arXiv:2607.24875, 2026. \href{https://doi.org/10.48550/arXiv.2607.24875}{\nolinkurl{doi:10.48550/arXiv.2607.24875}}.
\item[{\hfill[105]}] U.S. Securities and Exchange Commission. \href{https://www.sec.gov/rules/final/2022/34-96034.pdf}{Electronic Recordkeeping Requirements for Broker-Dealers, Security-Based Swap Dealers, and Major Security-Based Swap Participants}. Release 34-96034 (amendments to Rule 17a-4), 2022.
\item[{\hfill[106]}] Huisheng Wang, Zhuoshi Pan, Hangjing Zhang, Mingxiao Liu, Hanqing Gao, and H. Vicky Zhao. InvestAlign: Overcoming Data Scarcity in Aligning Large Language Models with Investor Decision-Making Processes under Herd Behavior. arXiv preprint arXiv:2507.06528, 2025. \href{https://doi.org/10.48550/arXiv.2507.06528}{\nolinkurl{doi:10.48550/arXiv.2507.06528}}.
\item[{\hfill[107]}] Mengxiao Wang and Nitesh Saxena. Trading Agents or Market Crashers? Dissecting Robustness and Security Failures in Academic Financial LLM Trading Schemes. arXiv preprint arXiv:2609.19705, 2026. \href{https://doi.org/10.48550/arXiv.2609.19705}{\nolinkurl{doi:10.48550/arXiv.2609.19705}}.
\item[{\hfill[108]}] Saizhuo Wang, Hang Yuan, Leon Zhou, Lionel M. Ni, Heung-Yeung Shum, and Jian Guo. Alpha-GPT: Human-AI Interactive Alpha Mining for Quantitative Investment. arXiv preprint arXiv:2308.00016, 2023. \href{https://doi.org/10.48550/arXiv.2308.00016}{\nolinkurl{doi:10.48550/arXiv.2308.00016}}.
\item[{\hfill[109]}] Yiqi Wang, Jiaqi Zhang, Zhangkai Wu, Taotao Cai, Zirui Liu, Zequn Sun, Mingkai Zheng, Manqing Dong, Yu-Xuan Qiu, Xuefei Yin, and Yanming Zhu. From Agent Traces to Trust: A Survey of Evidence Tracing and Execution Provenance in LLM Agents. arXiv preprint arXiv:2606.04990, 2026. \href{https://doi.org/10.48550/arXiv.2606.04990}{\nolinkurl{doi:10.48550/arXiv.2606.04990}}.
\item[{\hfill[110]}] Zexin Wang, Jingjing Li, Quan Zhou, Haotian Si, Yuanhao Liu, Jianhui Li, Gaogang Xie, Fei Sun, Dan Pei, and Changhua Pei. A Survey on AgentOps: Categorization, Challenges, and Future Directions. arXiv preprint arXiv:2508.02121, 2025. \href{https://doi.org/10.48550/arXiv.2508.02121}{\nolinkurl{doi:10.48550/arXiv.2508.02121}}.
\item[{\hfill[111]}] Yihan Xia, Panpan You, Taotao Wang, Fang Liu, Han Qi, Xiaoxiao Wu, and Shengli Zhang. Agentic Trading: When LLM Agents Meet Financial Markets. arXiv preprint arXiv:2605.19337, 2026. \href{https://doi.org/10.48550/arXiv.2605.19337}{\nolinkurl{doi:10.48550/arXiv.2605.19337}}.
\item[{\hfill[112]}] Yijia Xiao, Edward Sun, Di Luo, and Wei Wang. TradingAgents: Multi-Agents LLM Financial Trading Framework. arXiv preprint arXiv:2412.20138, 2024. \href{https://doi.org/10.48550/arXiv.2412.20138}{\nolinkurl{doi:10.48550/arXiv.2412.20138}}. Oral presentation at the Multi-Agent AI in the Real World workshop.
\item[{\hfill[113]}] Yijia Xiao, Edward Sun, Tong Chen, Fang Wu, Di Luo, and Wei Wang. Trading-R1: Financial Trading with LLM Reasoning via Reinforcement Learning. arXiv preprint arXiv:2509.11420, 2025. \href{https://doi.org/10.48550/arXiv.2509.11420}{\nolinkurl{doi:10.48550/arXiv.2509.11420}}.
\item[{\hfill[114]}] Fei Xiong, Xiang Zhang, Aosong Feng, Siqi Sun, and Chenyu You. QuantHarness: Price-Driven Multi-Agent LLMs for High-Frequency Trading. arXiv preprint arXiv:2509.09995, 2025. \href{https://doi.org/10.48550/arXiv.2509.09995}{\nolinkurl{doi:10.48550/arXiv.2509.09995}}. Earlier versions titled QuantAgent.
\item[{\hfill[115]}] Guojun Xiong, Zhiyang Deng, Keyi Wang, Yupeng Cao, Haohang Li, Yangyang Yu, Xueqing Peng, Mingquan Lin, Kaleb E. Smith, Xiao-Yang Liu Yanglet, Jimin Huang, Sophia Ananiadou, and Qianqian Xie. FLAG-Trader: Fusion LLM-Agent with Gradient-based Reinforcement Learning for Financial Trading. In Findings of the Association for Computational Linguistics: ACL 2025, Vienna, Austria, pages 13921–13934, 2025. \href{https://doi.org/10.18653/v1/2025.findings-acl.716}{\nolinkurl{doi:10.18653/v1/2025.findings-acl.716}}.
\item[{\hfill[116]}] Weicheng Xue. Representation Signatures and Risk-Feedback Alignment in LLM Trading Agents. arXiv preprint arXiv:2605.28850, 2026. \href{https://doi.org/10.48550/arXiv.2605.28850}{\nolinkurl{doi:10.48550/arXiv.2605.28850}}.
\item[{\hfill[117]}] Lewen Yan, Jilin Mei, Tianyi Zhou, Lige Huang, Jie Zhang, Dongrui Liu, and Jing Shao. TradeTrap: Are LLM-based Trading Agents Truly Reliable and Faithful? arXiv preprint arXiv:2512.02261, 2025. \href{https://doi.org/10.48550/arXiv.2512.02261}{\nolinkurl{doi:10.48550/arXiv.2512.02261}}.
\item[{\hfill[118]}] Yuzhe Yang, Yifei Zhang, Minghao Wu, Kaidi Zhang, Yunmiao Zhang, Honghai Yu, Yan Hu, and Benyou Wang. TwinMarket: A Scalable Behavioral and Social Simulation for Financial Markets. arXiv preprint arXiv:2502.01506, 2025. \href{https://doi.org/10.48550/arXiv.2502.01506}{\nolinkurl{doi:10.48550/arXiv.2502.01506}}.
\item[{\hfill[119]}] Junyi Yao, Zihao Zheng, and Baichuan Li. Execution Realism and Reproducibility in LLM-Based Trading Systems: A Systematic Scoping Review and Evidence Audit. arXiv preprint arXiv:2606.08285, 2026. \href{https://doi.org/10.48550/arXiv.2606.08285}{\nolinkurl{doi:10.48550/arXiv.2606.08285}}.
\item[{\hfill[120]}] Haofei Yu, Fenghai Li, and Jiaxuan You. LiveTradeBench: Seeking Real-World Alpha with Large Language Models. arXiv preprint arXiv:2511.03628, 2025. \href{https://doi.org/10.48550/arXiv.2511.03628}{\nolinkurl{doi:10.48550/arXiv.2511.03628}}.
\item[{\hfill[121]}] Yangyang Yu, Zhiyuan Yao, Haohang Li, Zhiyang Deng, Yuechen Jiang, Yupeng Cao, Zhi Chen, Jordan W. Suchow, Zhenyu Cui, Rong Liu, Zhaozhuo Xu, Denghui Zhang, Koduvayur Subbalakshmi, Guojun Xiong, Yueru He, Jimin Huang, Dong Li, and Qianqian Xie. FinCon: A Synthesized LLM Multi-Agent System with Conceptual Verbal Reinforcement for Enhanced Financial Decision Making. In Advances in Neural Information Processing Systems 37 (NeurIPS 2024), 2024. URL \href{https://proceedings.neurips.cc/paper/2024/hash/f7ae4fe91d96f50abc2211f09b6a7e49-Abstract-Conference.html}{\nolinkurl{https://proceedings.neurips.cc/paper/2024/hash/f7ae4fe91d96f50abc2211f09b6a7e49-Abstract-Conference.html}}. Also arXiv:2407.06567.
\item[{\hfill[122]}] Yangyang Yu, Haohang Li, Zhi Chen, Yuechen Jiang, Yang Li, Denghui Zhang, Rong Liu, Jordan W. Suchow, and Khaldoun Khashanah. FinMem: A Performance-Enhanced LLM Trading Agent with Layered Memory and Character Design. In Proceedings of the AAAI Symposium Series, 3(1):595–597, 2024. \href{https://doi.org/10.1609/aaaiss.v3i1.31290}{\nolinkurl{doi:10.1609/aaaiss.v3i1.31290}}.
\item[{\hfill[123]}] Hang Yuan, Saizhuo Wang, and Jian Guo. Alpha-GPT 2.0: Human-in-the-Loop AI for Quantitative Investment. arXiv preprint arXiv:2402.09746, 2024. \href{https://doi.org/10.48550/arXiv.2402.09746}{\nolinkurl{doi:10.48550/arXiv.2402.09746}}.
\item[{\hfill[124]}] Chong Zhang, Xinyi Liu, Zhongmou Zhang, Mingyu Jin, Lingyao Li, Zhenting Wang, Wenyue Hua, Dong Shu, Suiyuan Zhu, Xiaobo Jin, Sujian Li, Mengnan Du, and Yongfeng Zhang. When AI Meets Finance (StockAgent): Large Language Model-based Stock Trading in Simulated Real-world Environments. arXiv preprint arXiv:2407.18957, 2024. \href{https://doi.org/10.48550/arXiv.2407.18957}{\nolinkurl{doi:10.48550/arXiv.2407.18957}}.
\item[{\hfill[125]}] Huopu Zhang, Yanguang Liu, Miao Zhang, Zirui He, and Mengnan Du. SAE-FiRE: Enhancing Earnings Surprise Predictions Through Sparse Autoencoder Feature Selection. arXiv preprint arXiv:2505.14420, 2025. \href{https://doi.org/10.48550/arXiv.2505.14420}{\nolinkurl{doi:10.48550/arXiv.2505.14420}}.
\item[{\hfill[126]}] Shaokun Zhang, Ming Yin, Jieyu Zhang, Jiale Liu, Zhiguang Han, Jingyang Zhang, Beibin Li, Chi Wang, Huazheng Wang, Yiran Chen, and Qingyun Wu. Which Agent Causes Task Failures and When? On Automated Failure Attribution of LLM Multi-Agent Systems. In Proceedings of the 42nd International Conference on Machine Learning, volume 267 of Proceedings of Machine Learning Research, pages 76583–76599, 2025. URL \href{https://proceedings.mlr.press/v267/zhang25cq.html}{\nolinkurl{https://proceedings.mlr.press/v267/zhang25cq.html}}.
\item[{\hfill[127]}] Wentao Zhang, Lingxuan Zhao, Haochong Xia, Shuo Sun, Jiaze Sun, Molei Qin, Xinyi Li, Yuqing Zhao, Yilei Zhao, Xinyu Cai, Longtao Zheng, Xinrun Wang, and Bo An. A Multimodal Foundation Agent for Financial Trading: Tool-Augmented, Diversified, and Generalist. In Proceedings of the 30th ACM SIGKDD Conference on Knowledge Discovery and Data Mining (KDD '24), Barcelona, Spain, 2024. \href{https://doi.org/10.1145/3637528.3671801}{\nolinkurl{doi:10.1145/3637528.3671801}}.
\item[{\hfill[128]}] Tianjiao Zhao, Jingrao Lyu, Stokes Jones, Harrison Garber, Stefano Pasquali, and Dhagash Mehta. AlphaAgents: Large Language Model based Multi-Agents for Equity Portfolio Constructions. arXiv preprint arXiv:2508.11152, 2025. \href{https://doi.org/10.48550/arXiv.2508.11152}{\nolinkurl{doi:10.48550/arXiv.2508.11152}}.
\item[{\hfill[129]}] Yuxuan Zhao, Sijia Chen, and Ningxin Su. PortBench: A Correlation-Aware, Full-Pipeline Benchmark for LLM-Driven Portfolio Management. arXiv preprint arXiv:2605.27887, 2026. \href{https://doi.org/10.48550/arXiv.2605.27887}{\nolinkurl{doi:10.48550/arXiv.2605.27887}}.
\item[{\hfill[130]}] Yuxuan Zhu and Peng Pu. TelemetrySuffBench: Is Agent Telemetry Sufficient for Failure-Origin Diagnosis? arXiv preprint arXiv:2608.07899, 2026. \href{https://doi.org/10.48550/arXiv.2608.07899}{\nolinkurl{doi:10.48550/arXiv.2608.07899}}.
\end{list}
\endgroup

\clearpage
\appendix
\setcounter{figure}{0}\setcounter{table}{0}\renewcommand{\thefigure}{A\arabic{figure}}\renewcommand{\thetable}{A\arabic{table}}

\section{Record Model and Signal Definitions}

\subsection{Record Model}

To make the signals traceable across stages and across agents, FinAO assumes a common record model. Every record is a span, an event, a state snapshot or an evidence record. Records are joined by typed links such as \texttt{delegates\_to}, \texttt{uses\_context}, \texttt{derived\_from} and \texttt{validated\_by}, and each link is marked as observed, declared or inferred, following the evidence-quality grading. The links are what let a review follow a belief from one agent to another in a multi-agent desk [121], or an order back to the intent that authorized it [36]. In the running example, the stored belief carries a derived\_from link to the Day 1 headline, and the Day 3 and Day 5 orders each carry a uses\_context link to the belief. Figure A1 lists the four questions about an agent's behavior and shows what must be recorded to answer each one.

\begin{figure}[htbp]
\centering
\includegraphics[width=0.86\linewidth]{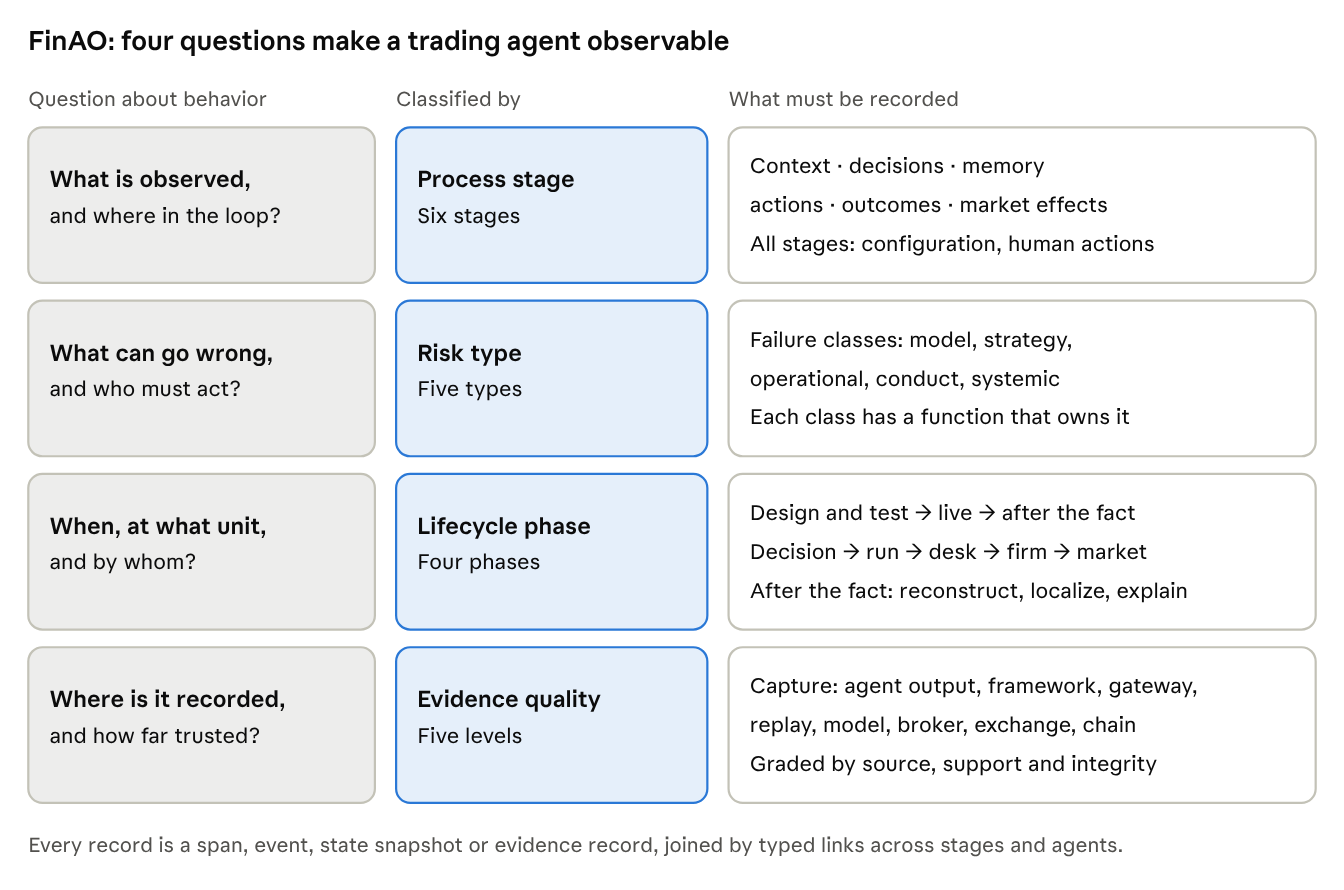}
\caption{FinAO as an observability taxonomy. Stage, risk type, lifecycle phase and evidence quality each answer one question about an agent's behavior and fix what must be recorded to answer it.}
\label{fig:figa1}
\end{figure}

\subsection{Metrics, Logs and Views by Stage}

Table A1 applies the taxonomy stage by stage: what to measure, what to log and trace, and what a person reviewing the agent should be able to read. The metrics are standard finance measures; the logs and traces follow the record model above.

\begingroup\footnotesize\setlength{\tabcolsep}{4pt}\renewcommand{\arraystretch}{1.18}
\begin{longtable}{>{\raggedright\arraybackslash}p{0.099\dimexpr\linewidth-10\tabcolsep\relax}>{\raggedright\arraybackslash}p{0.093\dimexpr\linewidth-10\tabcolsep\relax}>{\raggedright\arraybackslash}p{0.316\dimexpr\linewidth-10\tabcolsep\relax}>{\raggedright\arraybackslash}p{0.254\dimexpr\linewidth-10\tabcolsep\relax}>{\raggedright\arraybackslash}p{0.223\dimexpr\linewidth-10\tabcolsep\relax}}
\caption{Observability signals by investment-process stage.}\\
\toprule
\textbf{Stage} & \textbf{Observed target} & \textbf{Metrics} & \textbf{Logs and traces} & \textbf{Human-readable view} \\
\midrule
\endfirsthead
\toprule
\textbf{Stage} & \textbf{Observed target} & \textbf{Metrics} & \textbf{Logs and traces} & \textbf{Human-readable view} \\
\midrule
\endhead
\bottomrule
\endlastfoot
Data and information & Context; environment & Point-in-time compliance rate; gap between data date and knowledge cutoff; memorization score & Input snapshots with timestamps and sources; retrieval logs & Which inputs a decision used, and whether each was available at the time \\
Research and signal & Intent and decisions & Information coefficient; Brier score of stated confidence; rationale-to-trade consistency; style tilts & Prompts, outputs, debate transcripts and confidence, each as a span & A thesis whose every claim cites an input or record \\
Books and records & State and memory & Reconciliation breaks against broker or chain & Memory change log; belief lineage across agents & What the agent believed it held, and why that changed \\
Orders and execution & Actions and tool calls & Turnover; implementation shortfall; leverage and concentration; adherence to exit plan & Intent, authorization, order and fill, linked & Each order with its mandate check and authorization \\
Performance and risk attribution & Outcome evidence & Sharpe ratio before and after the cutoff; deflated Sharpe ratio; attribution by agent and signal & Fills, marks and profit and loss by signal & Which agent and signal produced the result \\
Market conduct and stability & External effects; coordination & LSV herding measure; cross-agent position correlation; price impact; supracompetitive margins & Inter-agent messages; trades by other agents & Who coordinated with whom, and the effect on prices \\
All stages & Configuration; runtime & Replay match rate; drift in model version and configuration; evaluation results over time & Model version, prompts, parameters, which model served each task, cost and latency & Exactly what was running when the decision was made \\
All stages & Human intervention & Override rate; time to approval & Approvals, overrides and escalations, each linked to its order & Who approved or overrode which order, and why \\
\end{longtable}\endgroup

\subsection{Signal Coding}

We coded each classified study for which signals of Table 1 its method records as evidence about agent behavior, from the same sources used for the FinAO coding. A study counts for a signal only when the signal is part of its evidence, not merely present in its setup. Naming the model does not count as recording its version, for example, but comparing behavior across training cutoffs does. Table A2 gives the rule for each signal. A single coder assigned these codes, and for many studies they rest on summaries rather than full papers, so they are provisional. Their pattern matches the evidence levels in Table 2.

\begingroup\footnotesize\setlength{\tabcolsep}{4pt}\renewcommand{\arraystretch}{1.18}
\begin{longtable}{>{\raggedright\arraybackslash}p{0.288\dimexpr\linewidth-4\tabcolsep\relax}>{\raggedright\arraybackslash}p{0.697\dimexpr\linewidth-4\tabcolsep\relax}}
\caption{Coding rule for each observability signal.}\\
\toprule
\textbf{Signal} & \textbf{What counts} \\
\midrule
\endfirsthead
\toprule
\textbf{Signal} & \textbf{What counts} \\
\midrule
\endhead
\bottomrule
\endlastfoot
Reasoning and confidence & Stated reasoning, debate or confidence is kept as evidence \\
Time-stamped inputs & What was available at each decision date is fixed or recorded \\
Model version and configuration & Behavior is attributed to a model version, training cutoff or configuration \\
Messages between agents & Inter-agent messages or coordination are recorded \\
Evaluation over time & Evaluation is repeated as outcomes arrive \\
State and memory & Changes to memory, beliefs, positions or the ledger are recorded \\
Order linked to intent & Each order is linked to the rationale, intent or authorization behind it \\
Human intervention & Human approvals, overrides, feedback or referrals are recorded \\
External records & Exchange, broker, on-chain or resolved-market records are used \\
Identical-input replay & Decisions are rerun on identical inputs to test repeatability \\
\end{longtable}\endgroup
\end{document}